\documentclass[12pt]{article}

\usepackage{hyperref,amssymb,amsfonts,amsmath,amsxtra,mathrsfs,latexsym,esint,epsfig,graphicx,subfigure,latexsym,esint,amsxtra,wrapfig,tocvsec2, makecell, multirow,diagbox}
\usepackage[mathscr]{eucal}

\hypersetup{
     colorlinks=true,
     linkcolor=black,
     citecolor=blue,
     urlcolor=blue
     }

\DeclareMathAlphabet{\mathpzc}{OT1}{pzc}{m}{it}

\allowdisplaybreaks[0]

\def \sn{\,\mbox{sn}}
\def \cn{\,\mbox{cn}}
\def \dn{\,\mbox{dn}}

\title{\boldmath Oscillatory states of quantum Kapitza pendulum}

\makeatletter
\renewcommand\@date{{%
  \vspace{0.7cm}%
  \large\centering
  \begin{tabular}{@{}c@{}}
    Wei He\footnote{weihephys@foxmail.com},\hspace{1 cm} Chang-Yong Liu\footnote{liuchangyong@nwsuaf.edu.cn}
  \end{tabular}%

  \vspace{0.7cm}

\begin{center}
\em{
  \textsuperscript{*}School of Physics and Astronomy, China West Normal University, Nanchong 637002, China\\
  \textsuperscript{$\dagger$}College of Science, Northwest A$\&$F University, Yangling 712100, China
  }
\end{center}
}}
\makeatother

\begin{document}
\baselineskip=18pt
\maxtocdepth{subsection}
\maketitle

\begin{center}
\textbf{Abstract}
\end{center}
\begin{quotation}
\noindent  We study quantum mechanics problem described by the Schr\"{o}dinger equation with Kapitza pendulum potential,
that is the asymmetric double-well potential on the circle.
For the oscillatory states spatially localize around the two stable saddle positions of the potential,
we obtain the perturbative eigenvalues and corresponding piecewise wavefunctions.
The spectrum is computed by extending the angle coordinate to the complex plane so that the quantization condition is formulated as contour integral along a path extending in the imaginary direction.
Quantum tunneling between the wells is computed.

\end{quotation}

\newpage

\pagenumbering{arabic}

\tableofcontents

\newpage

\section{Introduction}

The Kapitza pendulum is an interesting phenomenon in classical mechanics \cite{Kapitza1951, LandauLifschitzMechanics}.
A rigid pendulum with the suspension point subject to an oscillatory motion in the vertical direction, when the oscillatory frequency exceeds a critical value,
the effective potential of pendulum acquires an stable saddle position above the suspension point. Despite its simplicity as a mechanical problem,
the Kapitza pendulum implies a central concept of physics, that physics at enormously separated scales are decoupled,
the dynamics at higher frequency can be integrated out, physics at lower frequency is an effective theory encoding imprints from the higher frequency dynamics.

The Kapitza pendulum is an example of dynamical stabilization by periodic modulation, nowadays this mechanism is demonstrated in variety of ways.
There are experimental efforts to realise Kapitza pendulum at mesoscopic and microscopic systems.
For example, realising a micron size pendulum by optical tweezer method is reported in \cite{RichardsCubero2018},
creating an effective trap potential for ultracold neutral atoms using by laser beams with modulation frequency of order kilo hertz is reported in \cite{JiangOtt2021}.
As experimental realizations of microscopic Kapitza pendulum-like system enter the quantum regime, such time periodically modulated systems have to be studied using quantum mechanics.
Recently, a spin chain version of multiparticles Kapitza pendulum system was discussed in \cite{LeroseSilva2019}.
The quantum Kapitza pendulum has been actively studied because it is a representative example of Floquet engineering,
a general scheme to apply time periodic driving to quantum systems to achieve various novel properties such as new phases of synthetic quantum matter,
artificial gauge fields and materials with topologically non-trivial band structure \cite{BukovDAlessioPolkovnikov2014, OkaKitamura2019, WeitenbergSimonet2021, Rodriguez-VegaVoglFiete2021}.
This line of research provides tunable platforms to fabricate physical system with designed properties,
greatly enlarges accessible parameter space of some physics problems in the arena of condensed matter physics and quantum technology.

There are good reasons to pursue a better understanding of quantum mechanical problem with a Kapitza pendulum effective potential.
Primarily, it is a toy model for quantum particles in a periodic potential with two non-degenerate vacua.
Since we learn many basic properties of quantum theory from exactly or quasi-exactly solvable quantum mechanics models,
the quantum Kapitza pendulum, which is seemingly simple but far from trivial, could provide valuable lessons for quantum mechanics.
On the more practical side, the quantum mechanics potentials with two non-degenerate vacua are useful in the study of some physical systems.

In this work we consider some aspects of the quantum Kapitza pendulum,
in particular the perturbative spectrum. That means we simplify the problem by ignoring some common issues in experiments,
such as thermalization due to the interaction with surroundings, neither we consider theoretically more complicated cases such as multi-dimensional and multi-particle generalisations.
One may think the single particle perturbative quantum spectrum of the effective potential, which basically is the spectral problem of Schr\"{o}dinger equation with a three terms Hill's potential,
is very well understood as there are piles of literature on the physics and mathematics aspects of periodic differential equation.
This impression is only partially true, as far as we can say, even for a three terms Hill's potential the spectral property is rich,
and the Kapitza pendulum problem lies in a region in the parameter space not fully explored. Moreover,
we would study the spectrum of the quantum mechanical problem from a point of view unfamiliar in quantum mechanics,
namely using method of complex analysis by extending the angle coordinate to the complex domain.
The main tool we utilize is a generalised version of Floquet theorem for ordinary differential equation with periodic potential,
which has been discussed in the case with elliptic function potential that arises in some integrable quantum field theories,
the version applied in this paper is a degeneration to the case with real trigonometric function potential \cite{wh1108, wh1412, wh1608, wh1904}.

\section{Summary of  results}

The main results of this paper can be summarized as follows.

In section \ref{QuantumKapitzaPendulum}, after a brief introduction of the Kapitza effective potential,
we compute the eigensolution that describes rotating states with large kinetic energy, in the context of classical Floquet theory.
The rotating states are not localized, similar to the travelling states in extended space with periodic field, the spectral solution is well described by the Floquet-Bloch theorem.
On the contrary, as explained in \ref{PreliminaryDiscussionSpectrum}, the states with small kinetic energy oscillate at the bottom of the potential well, the quantum spectrum can not be treated similarly by classical Floquet theory.
Localized states in deep quantum wells are usually studied either by approximating the potential in the well region with a solvable potential or by the semiclassical WKB approximation \cite{Froman1972, Bhattacharya1985,  Song2015, MullerKirstenQUANT, GriffithsQUANT}.
In \cite{GolovinskiDubinkin2021}, the oscillatory states of quantum Kapitza potential is treated using the stationary perturbation method,
by approximating the potential around a well region with the harmonic oscillator potential.
In this approach, regarding the Floquet theorem, the oscillatory states and rotating states are not on the same footing.
As explained in our previous works \cite{wh1412, wh1608}, the oscillatory states of trigonometric function potentials do have the Floquet interpretation once the trigonometric functions are viewed as limit of the elliptic functions which are intrinsically complex.

In section \ref{OscillatoryStatesAtPhi0}, we compute the eigensolution for the oscillatory states localize in the well around $\phi_s=0$,
using the method of complex variable function.
This method is motivated by our study on the spectral solutions of Schr\"{o}dinger equation with elliptic function potentials,
it is noticed that by applying Floquet theorem to the imaginary period of elliptic function one computes the spectrum of small eigenvalue above the saddle position of the potential.
But the Kapitza effective potential is trigonometric, to utilize the power of Floquet theorem, we have to extend the coordinate to the complex plane.
The eigenvalue of oscillatory states are obtained by performing a large coupling perturbation computation, the Floquet theorem relates the quantum number to the contour integral of WKB integrand. Here the benefit of performing complex analysis comes, the residue theorem greatly simplifies the integrations. The series expansion of eigenvalue is given in (\ref{eigenvalue0}), the approximate wavefunctions applicable in the barrier region are given in (\ref{wavefunction0G}), and the wavefunctions applicable in the well region are given in (\ref{wavefunction0S}). By the approach we utilize, we can give a natural explanation to the integer value of the quantum number as result of contour integral of the integrand which has finite number of zeros if rewritten in the proper form of (\ref{integrandWithPoles}).
Moreover, the factor $1/2$ related to the zero point energy also arises naturally from the contour integral.
The oscillatory state can be interpreted as a Floquet-Bloch state associated to an imaginary period when the wavefunctions (\ref{wavefunction0G}) are rewritten in the canonical coordinate defined in (\ref{CanoniCoordDef}).

In section \ref{OscillatoryStatesAtPhiPi}, the eigensolution of oscillatory states localize in the well around $\phi_s=\pi$ is given.
The computation follows the same steps as that for the states localize in the $\phi_s=0$ well, but the eigensolutions are slightly different since the two wells are asymmetric.
The series expansion of eigenvalue is given in (\ref{eigenvaluePi}), the approximate wavefunctions for barrier and well regions are given in (\ref{wavefunctionPiG}) and (\ref{wavefunctionPiS}), respectively. The table \ref{SummaryPertSolu} in the last of this section summarize the perturbative eigensolutions obtained in sections \ref{OscillatoryStatesAtPhi0} and \ref{OscillatoryStatesAtPhiPi}.

Once an experiment is set up to measure the localized states of quantum Kapitza pendulum, the primary observable quantities are the energy levels and the localized wavepackets.
The spectrum discussed in section \ref{OscillatoryStatesAtPhi0} and \ref{OscillatoryStatesAtPhiPi} are energies above the perturbative vacua at the two saddle positions, we have to consider how much the nonperturbative tunneling affects the perturbative spectrum.
In section \ref{NonperturbativeTunneling} we use the WKB method to estimate the mixing of perturbative states and energy shift due to quantum tunneling. As far as the wells are deep and asymmetric, the mixing is exponentially suppressed, this is in accordance with the general results about quantum tunneling of double-wells potential \cite{Song2015, WeinerTse1981, Dekker1987, Song2008, Rastelli2012};
therefore, the eigensolutions given by (\ref{eigenvalue0})  (\ref{wavefunction0S}) and (\ref{eigenvaluePi}) (\ref{wavefunctionPiS}) are well preserved.
This also sets the limit for the results obtained in this paper, as the barrier is tuned shallow, quantum tunneling becomes strong and quantum states cease to localize in wells.

In \ref{Appendix1} some background materials on Floquet theorem for elliptic potentials are provided. In \ref{EigenvalueCurveMathieu}, using the Mathieu eigenvalue we explain the phenomenon of asymptotic coincidence of eigenvalues for oscillatory states with adjacent quantum numbers. Some useful formulae of parabolic cylinder functions are given in \ref{ParabolicCylinderFunction}.

Overall, the approach we utilize here not only introduces a method to compute the eigensolution of oscillatory states using complex contour integral,
it also provides a unified interpretation for both large energy rotating states and small energy oscillatory states in the context of Floquet-Bloch theory. The method can be applied to other similar quantum mechanical spectral problem with trigonometric function potentials.

\section{Quantum Kapitza pendulum}\label{QuantumKapitzaPendulum}

\subsection{The effective potential}

We first recall how to get the effective potential for a classical mechanical system with fast periodic driving.
Because the driving force contains slow and fast components, so do the response;
the long-term evolution of the system is given by average of the response over the fast modes.
It is true that the mean value of periodic driving force vanishes, but in the equation of motion the periodic driving is coupled to the slow motion component,
so that the long-time average leaves a non-vanishing term in the equation which contributes an extra potential term to the effective Hamiltonian,
thus slow component of the response is described by an effective potential that includes the imprint of fast periodic driving.
It is possible to change the parameters of driving force, typically the frequency, that the extra potential term could substantially change the nature of slow evolution,
this is how the Kapitza pendulum gains an inverted stable saddle position when the driving frequency is high enough.

For a quantum system with fast periodic driving of frequency $\omega$, one can also average the fast modes to obtain an effective Hamiltonian,
the process is performed in the quantum operator formalism \cite{GrozdanovRakovic1988, RahavGilaryFishman2003}. Once the long-time evolution operator $U(t_0+nT,t_0)$ is computed,
where $T=2\pi/\omega$ is the period of fast driving, the effective Hamiltonian $\mathcal{H}_F$ is defined by
\begin{equation}
U(t_0+nT,t_0)=\exp\left(-\frac{i \mathcal{H}_F}{\hbar}nT\right).
\end{equation}
The operator $U(t, t_0)$ satisfies the quantum evolution equation with time dependent Hamiltonian containing the periodic driving $f(t)$, $H(t)=H_0(\phi)+H_1(\phi)f(t)$,
where $\phi$ represents general coordinates of the system.
To solve the equation, $U(t, t_0)$ is assumed to take the form of exponential function with the exponent expanded as a series in $1/\omega$,
then the coefficient function at every order can be recursively solved. The effective Hamiltonian, hence the effective potential, is given by a Magnus series expanded in $1/\omega$.

\begin{figure}[hbt]
\begin{minipage}[t]{0.5\linewidth}
\centering
\includegraphics[width=5.5cm]{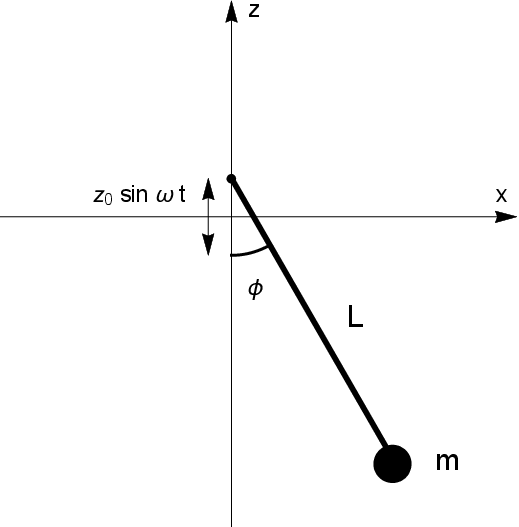}
\caption{Kapitza pendulum} \label{KapitzaPendulum}
\end{minipage}%
\begin{minipage}[t]{0.5\linewidth}
\centering
\includegraphics[width=5.5cm]{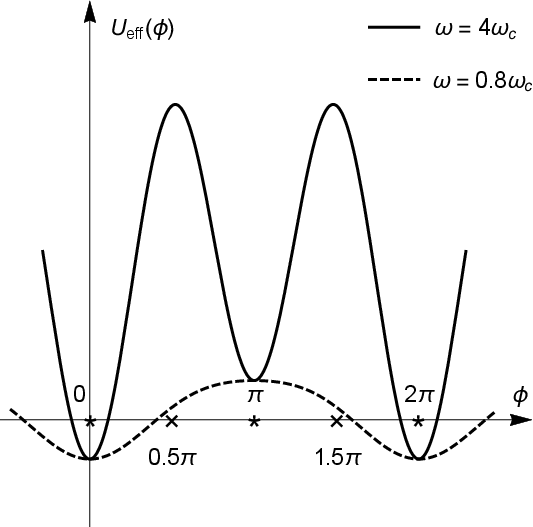}
\caption{The effective potential.} \label{KapitzaPendulumPotential}
\end{minipage}
\end{figure}

For a pendulum with mass $m$ and stick length $L$, when the suspension point is subjected to vertical vibration with frequency $\omega$ and amplitude $z_0$, see Fig. \ref{KapitzaPendulum},
the leading order effective potential is
\begin{equation}
U_{eff}(\phi)=-m\omega_0^2L^2\cos\phi+\frac{1}{4}m\omega^2 z_0^2\sin^2\phi,
\end{equation}
where $\omega_0=\sqrt{g/L}$ is the frequency of slow driving comes from gravity; and $\phi$ is the angle coordinate, therefore the potential is a Hill's potential of period $2\pi$.
The second term comes from averaging the fast component and its magnitude grows with frequency.
When the frequency $\omega$ is not high enough, the shape of the effective potential is dominated by the first term, the pendulum has one stable saddle at $\phi=0$;
when it exceeds a critical value, $\omega_c=(\sqrt{2}L/z_0)\omega_0$, the potential acquires an extra stable saddle at $\phi=\pi$.
There are two unstable saddle positions determined by the equation $\cos\phi=-2\omega_0^2L^2/\omega^2z_0^2$. These features are shown in Fig. \ref{KapitzaPendulumPotential}.
The quantum state of the pendulum satisfies the Schr\"{o}dinger equation
\begin{equation}
-\frac{\hbar^2}{2mL^2}\frac{\partial^2}{\partial\phi^2}\Psi(\phi,t)+U_{eff}(\phi)\Psi(\phi,t)=i\hbar\frac{\partial}{\partial t}\Psi(\phi,t).
\end{equation}
For a stationary state with energy $E$, set $\Psi(\phi,t)=\exp(-iEt/\hbar)\psi(\phi)$, then the equation becomes
\begin{equation}
-\frac{\hbar^2}{2mL^2}\psi^{\,\prime\prime}(\phi)+U_{eff}(\phi)\psi(\phi)=E\psi(\phi).
\end{equation}
the prime denotes the derivative w.r.t. the angle $\phi$.

In microscopic realization, the driving forces are often not gravity and external mechanical vibration force, maybe optical or otherwise, only the shape of effective potential is essential.
To have an equation applicable to general quantum systems, we write it in a dimensionless form as
\begin{equation}
\psi^{\,\prime\prime}(\phi)+[\mathscr{E}-u(\phi)]\psi(\phi)=0, \label{SchrodEqPhi}
\end{equation}
with potential
\begin{equation}
u(\phi)=-A\cos\phi+B\sin^2\phi, \label{effectivepotential}
\end{equation}
where
\begin{equation}
A=\frac{2m^2\omega_0^2L^4}{\hbar^2},\quad B=\frac{m^2\omega^2z_0^2L^2}{2\hbar^2},\quad \mathscr{E}=\frac{2mL^2E}{\hbar^2}.
\end{equation}
At the summit points $\phi_s$ determined by $\cos\phi_s=-A/2B$, the potential reaches the highest value $u(\phi_s)\approx B+A^2/4B$.
The equation is equivalent to the Whittaker-Hill equation,
\begin{equation}
\widetilde{\psi}^{\,\prime\prime}(x)+(\theta_0+\theta_1\cos 2x+\theta_2\cos 4x )\widetilde{\psi}(x)=0,
\end{equation}
with
\begin{equation}
\theta_1=\frac{8m^2\omega_0^2L^4}{\hbar^2},\quad \theta_2=\frac{m^2\omega^2z_0^2L^2}{\hbar^2},\quad \theta_0=\frac{8mL^2E}{\hbar^2}-\frac{m^2\omega^2z_0^2L^2}{\hbar^2}.
\end{equation}
To have a stable saddle at $\phi=\pi$, the ratio of the coefficients should satisfy $B/A> 0.5$, or $\theta_2/\theta_1>0.25$.
In atomic and molecular experimental setups, the frequency of driving could vary a wide band, so there ar plenty of possibilities to realise the potential with different shapes.
We are particularly interested in the case when $B$ is numerically large so that $B^{1/2}\gg 1$ and $B/A>0.5$ are both satisfied, then the potential has two deep wells to host local oscillatory states; these states are the quantum counterpart of classical pendulum vibrations.

The trigonometric series solutions of the Whittaker-Hill equation, called {\em paraboloidal wave functions},  has been studied by Ince, Arscott, etc, see \cite{Arscott1967, Arscott1964},
however, the solutions discussed there are suitable for the case with parameters $A<0$ and $B^{1/2}\ll |A|$, therefore not applicable to the present problem.

\subsection{Rotating states}\label{RotatingStates}

When the pendulum is very energetic, $\mathscr{E}\gg B$, the energy exceeds the potential barrier, the pendulum moves like a rotor, modulated by the effective potential.
The rotating state is a Floquet-Bloch state described by the Floquet solution of the equation (\ref{SchrodEqPhi}).
The general theory of periodic differential equation is well understood \cite{Arscott1964, mclachlan1947, Magnus-Winkler1966, Eastham1973}.
There are two independent solutions, they take a {\em quasi-periodic} form, $\psi_\pm(\phi)=\exp(\pm i\nu\phi)f_\pm(\phi)$, where $f_\pm(\phi)$ is a function of periodic $2\pi$,
then under the periodic translation $\psi_\pm(\phi+2\pi)=\exp(\pm i2\pi \nu)\psi_\pm(\phi)$.
The factor $\exp(\pm 2\pi i\nu)$ is the {\em characteristic multiplier}, the quantum number $\nu$ is the {\em characteristic exponent}.

In the following we show how the Floquet theorem can be used to compute the solution of rotating states, following a procedure explained in \cite{wh1412, wh1608}.
Assuming the eigenfunction in exponential form, $\psi(\phi)=\exp(\int^\phi v(\phi^{\,\prime})d\phi^{\,\prime})$, then the integrand $v(\phi)$ satisfies a nonlinear equation,
\begin{equation}
\partial_\phi v(\phi)+v^2(\phi)=u(\phi)-\mathscr{E},
\label{eq4integrand}
\end{equation}
in general it has no closed solution for the potential (\ref{effectivepotential}).
For a fast rotating state, the energy $\mathscr{E}$ can be used as the expansion parameter to find series solutions for the integrand,
there are ``$\pm$" sector solutions,
\begin{equation}
v_\pm(\phi)=\pm i\mathscr{E}^{1/2}\pm\frac{i(A\cos\phi-B\sin^2\phi)}{2\mathscr{E}^{1/2}}+\frac{B\sin2\phi+A\sin\phi}{4\mathcal{E}}+\cdots,
\label{integrandEexpansion}
\end{equation}
they are related by the interchange $\mathscr{E}^{1/2} \to -\mathscr{E}^{1/2}$.
The index is given by integration of $v_\pm(\phi)$ performed along the circle $\phi$,
\begin{equation}
\pm \nu=\frac{1}{2\pi i}\oint v_{\pm}(\phi^{\,\prime})d\phi^{\,\prime}.
\end{equation}
From the series expansion of $\nu(\mathscr{E})$, we obtain the eigenvalue $\mathscr{E}(\nu)$,
\begin{equation}
\mathscr{E}_r=\nu^2+\frac{B}{2}+\frac{B^2+2A^2}{32\nu^2}+\frac{2B^2-3A^2B+2A^2}{64\nu^4}+\cdots.
\end{equation}
The corresponding wavefunctions are obtained from indefinite integral of $v_\pm(\phi)$,
\begin{align}
\psi_{r\pm}(\phi)=&C_{r\pm}\exp(\pm i\nu\phi)\Big\{1\pm\frac{i}{8\nu}[B\sin2\phi+4A\sin\phi]+\frac{1}{256\nu^2}[B^2\cos4\phi+8AB\cos3\phi\nonumber\\
&-16(2B-A^2)\cos2\phi-8A(B+8)\cos\phi]+\cdots\Big\}.
\label{wavefunctionrotating}
\end{align}
The uniqueness of wavefunctions under translation $\phi\to\phi+2\pi$ require the index $\nu$ to take integer values.
Moreover, the eigenvalue is degenerate for $\pm\nu$, so $\nu$ takes values only in natural numbers.
The normalization constants are real and equal,
\begin{equation}
C_+=C_-=\sqrt{2\pi}\left[1-\frac{B}{8\nu^2}+\frac{B^2-8B+6A^2}{64\nu^4}+\mathcal{O}\left(\frac{B^3}{\nu^6}\right)\right],
\end{equation}
so that wavefunctions are complex conjugated, $\psi_+(\phi)=\overline{\psi_-(\phi)}$.
Since under the reflection $\phi\to -\phi$, the wavefunctions are related by $\psi_+(-\phi)=\psi_-(\phi)$,
one can construct linear combinations $\psi_C(\phi)=\psi_+(\phi)+\psi_-(\phi)$ and $\psi_S(\phi)=\psi_+(\phi)-\psi_-(\phi)$ with even and odd parities, respectively.
In the limit $A\to 0$, $\psi_C(\phi)$ reduces to the Mathieu $ce_\nu(\phi)$ function and $\psi_S(\phi)$ reduces to the $se_\nu(\phi)$ function.

\subsection{Preliminary discussion on oscillatory states}\label{PreliminaryDiscussionSpectrum}

In the case of large driving frequency, $B>A$, the effective potential has two wells at the saddle positions $\phi_s=0$ and $\phi_s=\pi$.
In particular, when $B$ is significantly larger than $A$, both potential wells are deep.
Near the bottom of the wells there are oscillatory modes, these states are spatially localized around the saddle positions.
Since an oscillatory state has a small excitation energy above the local vacuum, the potential term outplays the kinetic term in determining the behavior of wavefunction.
A consequence is that the integrand solution given by (\ref{integrandEexpansion}) is not applicable for oscillatory states, thus the argument based on classical Floquet theorem fails, too.

Quantum mechanically, the particle has probability to appear outside the well, the wavefunction spreads to the barrier region, even leaks to the other well with sizable amplitude.
As the potential varies substantially over $[0, 2\pi]$, the behavior of wavefunction differs significantly at different regions of the potential.
It is difficult to find a solution for $v(\phi)$, hence an expression of wavefunction $\psi(\phi)$, for an oscillatory state that can be applied over the whole angle;
instead, one can only find an approximation solution which is applicable in a particular region.
Therefore, at the moment the best we have is approximations of the true wavefunction at different regions and glue them together to get a global understanding.

Near the bottom, the potential $u(\phi)$ is approximately a harmonic potential, the discrepancy is represented by higher order corrections.
The eigenvalue can be computed by the stationary perturbation method \cite{GolovinskiDubinkin2021}.
Among others, a quantum number $n$ which takes value in natural numbers, is introduced by matching the leading order eigenvalue with harmonic oscillator eigenvalue.
The method works in practice, we nonetheless would ask the question: what is the quantization condition that assigns the integer quantum number, and can be used to derive the perturbative eigenvalue?

A quantization condition can be formulated in different ways.
For the harmonic oscillator problem, quantization condition arises from the requirement that the polynomial part of wavefunction must be truncated at the $n$-th order to ensure finiteness,
or it arises from joining semi-classical wavefunction in the common region which leads to the Bohr-Sommefeld quantization.
For the oscillatory states of quantum Kapitza problem, the potential (\ref{effectivepotential}) has a finite height and a compact angle coordinate, finiteness of wavefunction is not an issue;
moreover, the piecewise wavefunctions we compute later are already periodic prior to a quantization condition is applied,
so it is unlikely to get quantization condition from the uniqueness of wavefunction under translation $2\pi$ of the angle coordinate.
However, similar to the treatment of quantum mechanics with symmetric double-well potential \cite{Froman1972, Bhattacharya1985, MullerKirstenQUANT, GriffithsQUANT} or asymmetric double-well potential \cite{Song2015},
it is possible to derive a quantization condition of Bohr-Sommefeld type by enforcing proper matching condition on the WKB wavefunctions,  with the nonperturbative barrier tunneling effect included.
Therefore, it can be used to compute not only the perturbative eigenvalue but also the energy splitting caused by tunneling.

In this paper, we formulate the quantization condition of oscillatory states differently, as a contour integral in the complex plane.
The discussion relies on complexification of the quantum Kapitza pendulum, the coordinate $\phi$ is extended to the complex plane, while the parameters $A, B$ and $\mathscr{E}$ remain real.
Then it is possible to choose a proper contour $\mathcal{C}$ that under the translation of coordinate $\phi\to\phi+\mathcal{C}$ the wavefunction produces a phase given by integral of $v_\pm(\phi)$ along the contour,
where $v_\pm(\phi)$ are integrands solved from the equation (\ref{eq4integrand}) for small energy; if the contour circles simple poles of the integrand then a quantization condition arises naturally.

\section{Oscillatory states at the saddle position $\phi_s=0$}\label{OscillatoryStatesAtPhi0}

\subsection{Eigenvalue expansion and wavefunctions in the barrier}\label{WavefunctionInsideBarrier0}

In this section we study the quantum states that localize near the saddle $\phi_s=0$, later we describe the situation at the saddle $\phi_s=\pi$ which is similar.
As discussed in \ref{PreliminaryDiscussionSpectrum}, we are limited by analytical tool to obtain approximations of the wavefunction only applicable in some regions,
but these approximate wavefunctions are associated to the same perturbative eigenvalue.
Realising this difficulty, we shall first compute the eigenvalue using an very efficient method,
the wavefunction obtained is applicable in the barrier region around $\phi=\pi/2$;
then with the known eigenvalue it is easier to compute the wavefunction applicable in the nearby well region.

For the nonlinear equation (\ref{eq4integrand}), there is a series solution for large $B$, as explained in \cite{wh1412, wh1608}.
In the region where $\sin\phi$ is finite, the potential is much larger than the energy, $u(\phi)\sim\mathcal{O}(B)\gg \mathscr{E}$,
in this case we can use the potential as the large expansion quantity. The series solution of $v(\phi)$ expanded in $\pm\sqrt{u(\phi)}$,
or more conveniently expanded in $\pm B^{1/2}$, is given by
\begin{equation}
v_\pm(\phi)=\sum_{\ell=-1}^\infty\frac{v_\ell(\phi)}{(\pm B^{1/2})^\ell},
\label{integrandBexpansion}
\end{equation}
with $v_\ell(\phi)$ solved order by order,
\begin{align}
v_{-1}(\phi)=&-\sin\phi,\nonumber\\
v_0(\phi)=&-\frac{1}{2}\partial_\phi\ln\sin\phi,\nonumber\\
v_1(\phi)=&\frac{1}{2^3}\left(\frac{3}{\sin^3\phi}-\frac{1-4\mathscr{E}}{\sin\phi}\right)+\frac{A}{2}\partial_\phi\ln\sin\phi,\nonumber\\
v_2(\phi)=&\frac{1}{2^4}\partial_\phi\left(\frac{3}{\sin^4\phi}-\frac{1-4\mathscr{E}}{\sin^2\phi}\right)-\frac{A}{4}\left(\frac{2}{\sin^3\phi}-\frac{1}{\sin\phi}\right),\nonumber\\
v_3(\phi)=&\frac{1}{2^7}\left(\frac{297}{\sin^7\phi}-\frac{2(139-76\mathscr{E})}{\sin^5\phi}+\frac{(4\mathscr{E}+13)^2+16A^2-144}{\sin^3\phi}-\frac{16A^2}{\sin\phi}\right)\nonumber\\&
-\frac{A}{64}\partial_\phi\left(\frac{19}{\sin^4\phi}-\frac{2(5-4\mathscr{E})}{\sin^2\phi}\right), \cdots.
\label{integrands0}
\end{align}
With some terms properly written in total derivative form, the integrands $v_\ell(\phi)$ are expressed solely in the function $\sin\phi$.
The ``$\pm$" sector integrands lead to the corresponding eigenfunctions,
\begin{align}
\psi_{\pm}(\phi)=&C_\pm\exp\Big\{\pm B^{1/2}\cos\phi-\frac{1}{2}\ln\sin\phi\pm \frac{1}{2^4B^{1/2}}\Big[(8\mathscr{E}+1)\ln\tan\frac{\phi}{2}-\frac{3\cos\phi}{\sin^2\phi}\nonumber\\
& +8A\ln\sin\phi\Big]+\frac{1}{2^4B}\Big[\frac{3}{\sin^4\phi}-\frac{1-4\mathscr{E}-4A\cos\phi}{\sin^2\phi}\Big]+\cdots \Big\}.
\label{wavefunction0Gpre}
\end{align}
The integrands have a singularity at $\phi=0$, the wavefunctions derived from them are valid only in the barrier region away from the origin.
Because of this, the constants $C_\pm$ are not determined by normalization but by matching condition with wavefunction in the well region;
it is the $\phi$-dependent phases in (\ref{wavefunction0Gpre}) give useful information.
Only $\psi_+(\phi)$ decreases in the classical forbidden region away from the minima at $\phi_s=0$, therefore, it is the physically allowed wavefunction for a confined state.
Despite the procedure gives only the exponentially suppressed tail part of the wavefunction, it computes the eigenvalue efficiently, as we show in a moment.

Then we need a quantization scheme for the oscillatory modes, first of all, to get the relation between eigenvalue and quantum number.
The wavefunctions (\ref{wavefunction0Gpre}) do not take the form of quasi-periodic function, it is in fact formally periodic $2\pi$,
therefore no characteristic multiplier is produced under the translation $\phi\to\phi+2\pi$.
At the first glance, the Floquet theorem as explained in section \ref{RotatingStates} could not apply to the oscillatory states.
There is a way out, we shall extend the coordinate to the complex domain, then it is possible to choose a contour along which the integral of $v_\pm(\phi)$ equals an integer,
thus we have a quantization condition like the case for rotating states.

\begin{figure}[hbt]
\begin{minipage}[t]{0.5\linewidth}
\centering
\includegraphics[width=6cm]{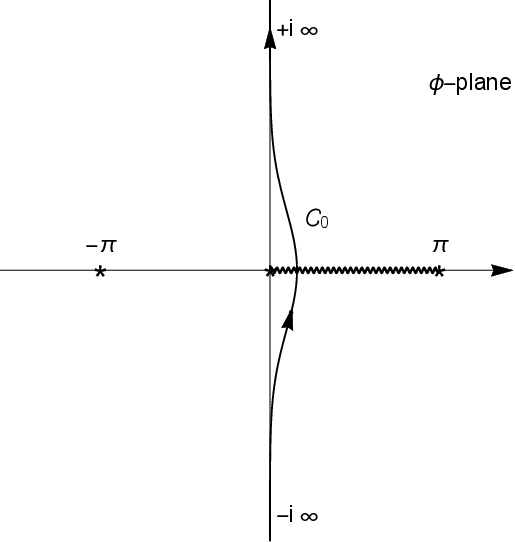}
\caption{Contour $\mathcal{C}_{0}$ and the branch cut\protect\\ for $\ln\sin\phi$ at $\phi_s=0$.} \label{PeriodContour4Phi0}
\end{minipage}%
\begin{minipage}[t]{0.5\linewidth}
\centering
\includegraphics[width=6cm]{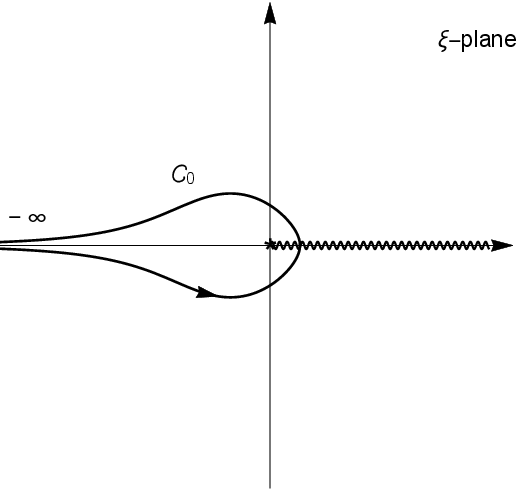}
\caption{Contour $\mathcal{C}_{0}$ in the $\xi$-coordinate and branch cut of $\ln\xi$.} \label{PeriodContour4Xi0}
\end{minipage}
\end{figure}

We extend the angle coordinate to the complex plane $\phi=x+iy$ and perform complex function computation.
The Floquet characteristic exponent, now denoted by $\mu$, is related to the monodromy of wavefunction translating along the contour $\mathcal{C}_{0}$, using the sector $v_+(\phi)$ it is computed by
\begin{equation}
\mu=\frac{1}{i\pi}\int_{\mathcal{C}_{0}} v_+(\phi)d\phi. \label{periodintegral0p}
\end{equation}
The contour $\mathcal{C}_{0}$ goes from $-i\infty$ to $+i\infty$ and passes close by the singular point $\phi=0$ counter-clockwise, as shown in Fig. \ref{PeriodContour4Phi0}.
In the asymptotic infinity region of the contour $x\to 0$,  $y\to \mp\infty$, and $\sin\phi= i\sinh y\to\mp i\infty$, $\cos\phi= \cosh y\to\infty$;
therefor, in the imaginary direction $\psi_-(\phi)$ remains finite but $\psi_+(\phi)$ diverges.
Integration of $v_0(\phi)$ and $v_1(\phi)$ results the logarithm term $\ln\sin\phi$ which has a branch cut from $\phi=0$ to $\phi=\pi$,
the contour bends to the right to cross the branch cut.
If we have arranged the branch cut from $\phi=-\pi$ to $\phi=0$, near the singular points the contours should bend to the opposite side to cross the branch cut.

The contour as described above is inspired by study on the Schr\"{o}dinger equation with elliptic function potential in previous papers \cite{wh1412, wh1608}.
In that case, the spectral problem is intrinsically complex nature because the coordinate and parameters are all complex variables, the contour is an imaginary period of elliptic function;
therefore, a contour in the imaginary direction is very natural and well defined.
For the problem at hand, the potential is a trigonometric function with real angle coordinate.
As explained in Appendix \ref{Appendix1}, taking a degenerate limit, elliptic function reduces to trigonometric function; however, the imaginary period becomes infinitely extended.
In this sense, the contour $\mathcal{C}_0$ is an imaginary period of infinite length, but the periodic nature is totally lost in trigonometric function.

The contour integrals (\ref{periodintegral0p}) can be performed according to the prescription described in \cite{wh1412, wh1608}.
Transforming the coordinate to $\xi=\sin^2\phi$, the contours $\mathcal{C}_{0}$ is mapped to a circle in the $\xi$-plane\footnote{Contours in the $\phi$-plane and in the $\xi$-plane are related by the map $\xi(\phi)$,
but branch cuts are not, they are respectively determined by the logarithm functions $\frac{1}{2}\ln\sin\phi$ and $\frac{1}{2}\ln\xi$ in the exponent of wavefunction.},
also denoted by $\mathcal{C}_{0}$, stretched along the real axis to infinity on the left side, as shown in Fig. \ref{PeriodContour4Xi0}.
The integration acquires contribution from residues of terms with simple poles at $\xi=0$, the following integrations gives nonzero results,
\begin{align}
& \int_{\mathcal{C}_{0}}\partial_\phi\ln\sin\phi\, d\phi=\frac{1}{2}\oint_{\mathcal{C}_{0}}\frac{d\xi}{\xi}=i\pi,\nonumber\\
& \int_{\mathcal{C}_{0}}\frac{d\phi}{\sin^{2n+1}\phi}=\frac{1}{2}\oint_{\mathcal{C}_{0}}\frac{d\xi}{\xi^{n+1}\sqrt{1-\xi}}=i\pi\frac{(2n-1)!!}{(2n)!!},
\label{integralformulae}
\end{align}
for $n\geqslant 0$, and $(2n)!!$ etc. are the double factorial of integers; all other integration vanish.
Then formula (\ref{periodintegral0p}) leads to a large-$B$ series expansion for $\mu(\mathscr{E})$;
the inverse function is the dispersion relation $\mathscr{E}(\mu)$.
Using the integrand $v_+(\phi)$, the following eigenvalue expansion is obtained,
\begin{equation}
\mathscr{E}_0=-A+2B^{1/2}\tilde{\mu}-\frac{1}{8}(4\tilde{\mu}^2+1)-\frac{1}{32B^{1/2}}(4\tilde{\mu}^3+3\tilde{\mu}-16A\tilde{\mu})+\cdots,
\label{eigenvalue0}
\end{equation}
where
\begin{equation}
\tilde{\mu}=\mu+\frac{1}{2}.
\end{equation}

For the sector $v_-(\phi)$, notice the relation
\begin{equation}
v_-(\phi)=-v_+(\phi)+2\sum_{\ell=0}^{\infty}\frac{v_{2\ell}(\phi)}{B^{\ell}},
\end{equation}
and the contour integral of the even integrands are\footnote{In references \cite{wh1412, wh1608, wh1904}, the contour integral of total derivative term $\partial_\phi\ln\sin\phi$ from $v_0(\phi)$ and $v_1(\phi)$ is claimed to vanish. The claim is contravened by a more careful consideration. The function $\ln\sin\phi$ has a branch cut between $\phi=0$ and $\phi=\pi$, therefore, the contour integral of $\partial_\phi\ln\sin\phi$ is $i\pi$. To put it in another way, because $\partial_\phi\ln\sin\phi=\cot\phi$ has a simple pole at $\phi=0$, it contributes a residue $i\pi$ to the half-circle integration. The nonzero contribution from integral of $v_0(\phi)$ is responsible for the factor $1/2$ of zero point energy. All other total derivative terms $\partial_\phi\sin^{-2m}\phi$ with $m\geqslant 1$ are not affected by the branch cut since they have poles of higher order. The solutions presented in references \cite{wh1412, wh1608, wh1904} remain correct with the characteristic exponent $\mu$ substituted by $\tilde{\mu}$.}
\begin{equation}
\int_{\mathcal{C}_{0}}v_0(\phi)d\phi=-\frac{1}{2}i\pi,\qquad \int_{\mathcal{C}_{0}}v_{2\ell}(\phi)d\phi=0 \quad \mbox{for}\quad \ell\geqslant 1,
\end{equation}
then the integral of the sector $v_-(\phi)$ satisfies
\begin{equation}
-(\mu+1)=\frac{1}{i\pi}\int_{\mathcal{C}_{0}} v_-(\phi)d\phi. \label{periodintegral0n}
\end{equation}
The relation (\ref{periodintegral0n}) leads to the same eigenvalue (\ref{eigenvalue0}); in fact, the expression (\ref{eigenvalue0}) is invariant under the simultaneous inversions $B^{1/2}\to -B^{1/2}$ and  $\tilde{\mu}\to -\tilde{\mu}$, i.e. $\mu\to -(\mu+1)$.

\begin{figure}
\vspace{-20pt}
\begin{center}
\includegraphics[width=8cm]{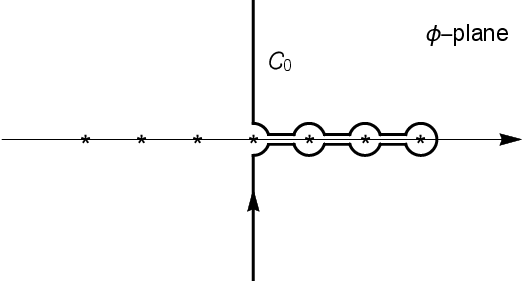}
\end{center}
\vspace{-10pt}
\caption{The contour $\mathcal{C}_0$ encircling half number of nodes, for the case $\mu=7$.} \label{NodesOfWavefunction}
\vspace{-5pt}
\end{figure}

Now we explain the reason for the index $\mu$ to take value in natural numbers.
The integrand $v_+(\phi)$ leads to the correct wavefunction $\psi_+(\phi)$ used to join with wavefunction in the classical allowed region.
In section \ref{WavefunctionInsideWell0} we show the wavefunction in the classical allowed region is approximately the harmonic oscillator wavefunction,
therefore there are simple zeros of the wavefunction distributed around $\phi_s=0$.
When the potential well is deep, the wavefunction is exponentially small in the other well, so all zeros are in the region around $\phi_s=0$.
A simple zero of the wavefunction, also called a node, corresponds to a simple pole of the integrand by the relation $v(\phi)=\partial_\phi\ln\psi(\phi)$.
This fact suggests the integrand can be written in another form  in the classically allowed region,
\begin{equation}
v_+(\phi)=\sum_{k=1}^{\mu}\frac{1}{\phi-\phi_k}+\mbox{regular\; terms},
\label{integrandWithPoles}
\end{equation}
where $\mu\geqslant 0$, and $\phi_k$ are the real simple zeros of wavefunction, situated symmetrically with respect to $\phi=0$.
The classical allowed region becomes narrow as the potential well gets deep, the zeros $\phi_k$ are squeezed into a small segment around the saddle position.
Then the contour $\mathcal{C}_0$ passing close by the saddle position in Fig. \ref{PeriodContour4Phi0} can be deformed to encircle {\em half} of the zeros,
as shown in Fig. \ref{NodesOfWavefunction}; thus the contour integral for $v_+(\phi)$ equals $\mu\pi i$ as in the definition (\ref{periodintegral0p}), through residue theorem.
This argument does not apply to higher level excitations with the eigenvalue comparable to the barrier height, because in that case there are nodes distributed in the regions of nearby barriers and the other well.
Indeed, for confined states in potential wells, the principle quantum number appearing in quantization condition is interpreted as counting the zeros of wavefunction \cite{MaXu2005A, MaXu2005B}.
For the Mathieu function of integer order, the index $\mu$ counts the number of real simple zeros of $se_\mu(\phi)$ and $ce_\mu(\phi)$ in $0\leqslant \phi<\pi$ \cite{Arscott1964, mclachlan1947};
here the index also counts the number of simple zeros of the wavefunction despite that the integrand we actually use to compute, given by (\ref{integrandBexpansion}) and (\ref{integrands0}),
obscures the pole structure.

In the argument given above, we do not consider poles associated with purely imaginary zeros.
There are infinite number of imaginary zeros for Mathieu functions $se_\mu(\phi)$ and $ce_\mu(\phi)$,
it is possible the wavefunctions of equation (\ref{SchrodEqPhi}) also have imaginary zeros.
Our computation indicates the contour integrals (\ref{periodintegral0p}) and (\ref{periodintegral0n}) should not be affected by purely imaginary zeros, even if they exist.
A possible reason is the following, only when the wavefunction contains a polynomial factor of finite order, like the harmonic oscillator wavefunction contains a Hermite polynomial,
which gives the singular part of  the integrand with finite number of poles as in  (\ref{integrandWithPoles}), and counting the number of simple zeros works as explained;
but in the imaginary direction, the simple zeros of wavefunction arise from functions of hyperbolic type which give the regular part of the integrand,
hence the imaginary zeros of that kind can not be seen in the expression (\ref{integrandWithPoles}). This issue needs further study.

Therefore, the eigenvalue given by (\ref{eigenvalue0}) is indeed the harmonic oscillator energy plus higher order perturbations,
$\mathscr{E}_0\approx 2(B+A/2)^{1/2}\tilde{\mu}\approx 2B^{1/2}(n+1/2)$, bounded from below as expected.
The dimensionless angular frequency is $\Omega_0=2(B+A/2)^{1/2}$; the factor $1/2$ is related to the zero point energy. When the depth of the well satisfies $B\gg B^{1/2}\tilde{\mu}$, that is $\tilde{\mu}\ll B^{1/2}$,
we can treat states with small $\tilde{\mu}$ as confined in the well,
and the expression (\ref{eigenvalue0}) gives a good approximation for the eigenvalue.

\begin{figure}
\vspace{-20pt}
\begin{center}
\includegraphics[width=5cm]{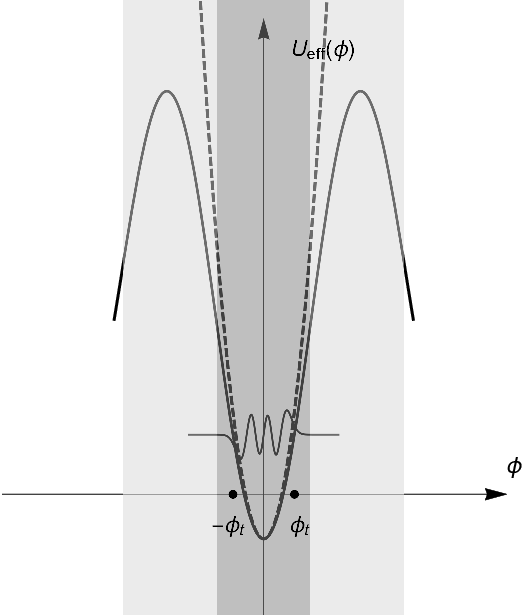}
\end{center}
\vspace{-10pt}
\caption{Potential around $\phi_s=0$, dark color covers the well region, light color covers the barrier regions.}
\label{PotentialAroundOrigin}
\vspace{-5pt}
\end{figure}

Substituting $\mathscr{E}(\tilde{\mu})$ into formula (\ref{wavefunction0Gpre}), the unnormalized eigenfunctions become
\begin{align}
\psi_{0\pm}(\phi)=&C_{0\pm}\exp\Big\{\pm B^{1/2}\cos\phi-\frac{1}{2}\ln\sin\phi\pm\tilde{\mu}\ln\tan\frac{\phi}{2}\pm\frac{1}{2^4B^{1/2}}\Big[\frac{\pm8\tilde{\mu}-(3+4\tilde{\mu}^2)\cos\phi}{\sin^2\phi} \nonumber\\ &+8A\ln\cos^2\frac{\phi}{2}\Big]+\frac{1}{2^6B}\Big[\frac{12+32\tilde{\mu}^2\mp(38\tilde{\mu}+8\tilde{\mu}^3)\cos\phi}{\sin^4\phi} \nonumber\\ &-\frac{(3+4\tilde{\mu}^2)(\pm\tilde{\mu}\cos\phi+2)+32A(1\pm\tilde{\mu})\sin^2\frac{\phi}{2}}{\sin^2\phi}\Big]+\cdots \Big\}.
\label{wavefunction0G}
\end{align}
The wavefunctions in (\ref{wavefunction0G}) are valid in the region of real slice of $\phi$ with
\begin{equation}
\sin^2\phi> \frac{\tilde{\mu}}{B^{1/2}}.
\end{equation}
The classical turning points $\pm\phi_t$ are determined by the equation $u(\phi_t)=\mathscr{E}$, for small $\tilde{\mu}$, that gives $\sin^2\phi_t\approx 2\tilde{\mu}/B^{1/2}$.
Therefore, the wavefunction $\psi_+(\phi)$ provides a good description in the barrier regions outside the turning points centered around $\phi=\pm \pi/2$, see Fig. \ref{PotentialAroundOrigin}.

\subsection{Interpreting the solution as Floquet-Bloch state}\label{InterpretingSolutionAsFloquet-Bloch}

\begin{figure}
\vspace{-20pt}
\begin{center}
\includegraphics[width=8cm]{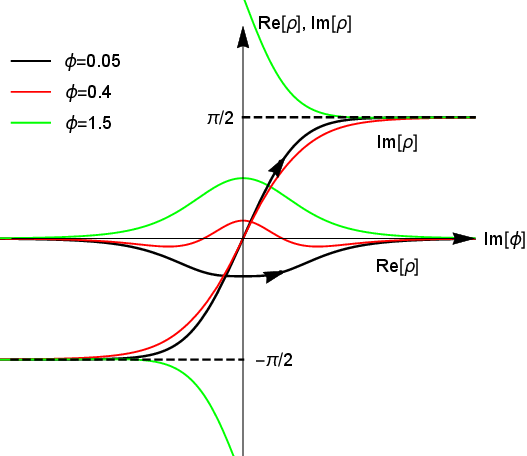}
\end{center}
\vspace{-10pt}
\caption{As $\phi$ translates by $\mathcal{C}_0$, the canonical coordinate $\rho$ changes $i\pi$.} \label{CanonicalCoordinateRho}
\vspace{-5pt}
\end{figure}

The wavefunction (\ref{wavefunction0G}) are Floquet solutions associated to the imaginary contour $\mathcal{C}_0$ in the following sense.
The exponent of the wavefunctions contains logarithm terms of order $\mathcal{O}(B^0)$, functions $\ln\sin\phi$ and $\ln\tan(\phi/2)$, which have a branch cut from $\phi=0$ to $\phi=\pi$.
When the angle $\phi$ shifts by the contour, the wavefunctions obtain a phase, the characteristic multiplier,
\begin{equation}
\psi_{\pm}(\phi+\mathcal{C}_0)\equiv\exp\left(\int_{\mathcal{C}_0}v_\pm(\phi)d\phi\right)\psi_{\pm}(\phi)=\left( \begin{matrix}
e^{i\pi\mu}  \cr e^{-i\pi(\mu+1)}
\end{matrix}\right)\psi_{\pm}(\phi).
\label{monodromyC0}
\end{equation}
In fact, similar to the study in \cite{wh1904}, one can define the canonical coordinate $\rho$ by
\begin{equation}
\ln\tan\frac{\phi}{2}=\rho,
\label{CanoniCoordDef}
\end{equation}
then the translation $\phi\to\phi+\mathcal{C}_{0}$ in the imaginary direction is equivalent to a finite coordinate shift $\rho\to\rho+i\pi$.
As shown in Fig. \ref{CanonicalCoordinateRho}, the evolution trajectories of $\rho(\phi)$ along the contour $\mathcal{C}_{0}$ for different value $\phi$ divide in the region around the origin,
but the overall change is always $\Delta\mbox{Re}\,\rho=0, \Delta\mbox{Im}\;\rho=i\pi$.

Using the following relations,
\begin{equation}
\sin^2\frac{\phi}{2}=\frac{e^{\rho}}{2\cosh\rho},\quad \cos^2\frac{\phi}{2}=\frac{e^{-\rho}}{2\cosh\rho},\quad \sin^2\phi=\frac{1}{\cosh^2\rho},\quad \cos\phi=-\tanh\rho,
\end{equation}
where all the functions of $\rho$ on the right side are periodic $i\pi$, the wavefunctions can be rewritten as quasi-periodic functions,
\begin{equation}
\psi_\pm(\rho)=\sqrt{\cosh\rho}\exp(\pm\widetilde{\mu}\rho)g_\pm(\rho),
\end{equation}
where $g_\pm(\rho)$ are functions of periodic $i\pi$, and the monodromy relation (\ref{monodromyC0}) becomes
\begin{equation}
\psi_\pm(\rho+i\pi)=\sqrt{-1}\exp(\pm i\pi\widetilde{\mu})\psi_\pm(\rho)=\left( \begin{matrix}
e^{i\pi\mu}  \cr e^{-i\pi(\mu+1)}
\end{matrix}\right)\psi_\pm(\rho),
\label{monodromyC0canoni}
\end{equation}
where we have chosen $\sqrt{-1}=\exp(-i\pi/2)$. Therefore, using the canonical coordinate $\rho$ the wavefunctions are expressed as Floquet function with an imaginary period;
of course they are valid only in the classically forbidden region around $\rho=0$.

The monodromy relation (\ref{monodromyC0}) indicates a subtle point that, although the wavefunction $\psi_+(\phi)$ acquires an index $\mu$ while the wavefunction $\psi_-(\phi)$ acquires an index $\mu+1$,
they are associated with the same perturbative eigenvalue. This is due to the presence of $1/2$ associated to ``zero point energy".
In contrast, wavefunctions for the large energy solutions (\ref{wavefunctionrotating}) discussed in section \ref{RotatingStates} acquire the same index.
This is consistent with the behavior of the eigenvalue for Mathieu equation when the coupling of potential varies from weak to strong, see Appendix \ref{EigenvalueCurveMathieu}.

Parity of the wavefunctions (\ref{wavefunction0G}) depends on the value of $\mu$,
\begin{equation}
\psi_{\pm}(-\phi)=\exp[i\pi(-1/2\pm\tilde{\mu})]\psi_{\pm}(\phi)=\pm(-1)^\mu\psi_{\pm}(\phi),
\end{equation}
which takes values $\pm 1$ when $\mu$ is an integer, $\psi_+(\phi)$ and $\psi_-(\phi)$ have opposite parities. The phase of parity comes from the logarithm terms.

The wavefunctions (\ref{wavefunction0G}) are similar to solutions of the Mathieu equation,
the expansion of Mathieu functions studied by Ince \cite{Ince1926second} and by Goldstein \cite{Goldstein1929},
see also Dingle and M\"{u}ller \cite{Dingle-Muller1962}, $\S 28.8$ of \cite{NISTDLMF}.
The Mathieu function $ce_\mu(\phi)$ and $se_{\mu+1}(\phi)$ are obtained from linear combinations $\psi_+(\pi/2-\phi)+\psi_-(\pi/2-\phi)$ and $\psi_+(\pi/2-\phi)-\psi_-(\pi/2-\phi)$ in the limit $A\to 0$, respectively\footnote{Mathieu function $ce_\mu(\phi)$ is parity even and $se_{\mu+1}(\phi)$ is parity odd;
in the relation given in this paragraph, because the argument is shifted by $\pi/2$,
the parity of $\psi_\pm(\phi)$ has nothing to do with the parity of Mathieu functions.}.

\subsection{Wavefunctions inside the potential well}\label{WavefunctionInsideWell0}

The approximate wavefunctions (\ref{wavefunction0G}) are applicable in the regions around $\phi=\pi/2$ and $\phi=3\pi/2$.
In the region inside the potential well around the saddle position $\phi=0$, since $\sin^2\phi\lesssim \tilde{\mu}/B^{1/2}$,
the potential and the energy are of the same order, $u(\phi)\sim \mathscr{E}\sim\mathcal{O}(B^{1/2}\tilde{\mu})$.
We need to find a series solution $v(\phi)$ of the equation (\ref{eq4integrand}) with large $B$, which is applicable in this region.
For the Mathieu function, the solution valid around $\phi=0$ is expanded in the parabolic cylinder function, studied by Sips \cite{Sips1949, Sips1959, Sips1965} and by Meixner \cite{Meixner1948, Meixner-Schafke1954},
see also \cite{Dingle-Muller1962} and $\S$28.8 of \cite{NISTDLMF}.
The reason is that near the saddle position $\phi_s=0$, the effective potential approximates the harmonic oscillator potential, so the wavefunction should approximates the parabolic cylinder function.

For the potential (\ref{effectivepotential}) of quantum Kapitza pendulum with large $B$,
the dominant term $B\sin^2\phi$ is a Mathieu potential with period $\pi$, there is a perturbative term $-A\cos\phi$.
The presence of $-A\cos\phi$ term does not change too much the structure of Sips-Meixner expansion of Mathieu function because, after all,
$-A\cos\phi$ is  a Mathieu potential with period $2\pi$.
In fact, the wavefunction of quantum Kapitza pendulum is also expanded in the parabolic cylinder function, as we show below.

Introducing the variable $z=\sqrt{2}B^{1/4}\sin\phi$, the Schr\"{o}dinger equation (\ref{SchrodEqPhi}) becomes
\begin{equation}
(2B^{1/2}-z^2)\psi^{\,\prime\prime}(z)-z\psi^{\;\prime}(z)+\left(\mathscr{E}+A\sqrt{1-\frac{z^2}{2B^{1/2}}}-\frac{1}{2}B^{1/2}z^2\right)\psi(z)=0.
\label{SchrodEqZ}
\end{equation}
In the region $|\sin\phi|\lesssim \tilde{\mu}^{1/2}/B^{1/4}$, and when the quantum number $\mu$ is not too large,
we have $z^2/2B^{1/2}<1$, the square root $\sqrt{1-z^2/2B^{1/2}}$ can be expanded as a series of large $B$.
Moreover, notice that despite the wavefunction is approximated by piecewise functions, the corresponding eigenvalue is unique,
so we can take the advantage of knowing the large-$B$ series of eigenvalue $\mathscr{E}$ given in (\ref{eigenvalue0}).
Then it is natural to expand the wavefunction in $B$ and assume the coefficient functions to take the form of Sips-Meixner series expansion.
There are two kinds of solutions,
\begin{align}
\psi_{C}(z)&=C_0 \sum_{\ell=0}^\infty\frac{\psi_{C}^{(\ell)}(z)}{(2B^{1/2})^\ell},\nonumber\\
\psi_{S}(z)&=S_0 \cos\phi\sum_{\ell=0}^\infty\frac{\psi_{S}^{(\ell)}(z)}{(2B^{1/2})^\ell},
\label{wavefunction0S}
\end{align}
with components $\psi_{C}^{(\ell)}(z)$ and $\psi_{S}^{(\ell)}(z)$ given by linear combination of parabolic cylinder function $D_n(z)$,
\begin{align}
\psi_{C}^{(\ell)}(z)&=\sum_{m=-2\ell}^{2\ell}C_{\ell,2m}D_{\mu+2m}(z),\nonumber\\
\psi_{S}^{(\ell)}(z)&=\sum_{m=-2\ell}^{2\ell}S_{\ell,2m}D_{\mu+2m}(z),
\label{PsiCSExpansion}
\end{align}
where $\mu$ takes value in natural number.
The constant coefficients $C_{\ell,2m}\equiv C_{\ell,2m}(\mu, A)$ and $S_{\ell,2m}\equiv S_{\ell,2m}(\mu, A)$ are polynomials of $\mu$ and $A$.

The equation (\ref{SchrodEqZ}) itself is a large-$B$ series, to the leading order it is an equation of the harmonic oscillator, the solution is $\psi^{(0)}(z)= D_\mu(z)$.
The equation also contains correction terms $z^2\psi^{\,\prime\prime}(z), z\psi^{\,\prime}(z)$ and $z^{2\ell}\psi(z)$ at order $B^{-\ell/2}$ for $\ell\geqslant 1$.
Starting from the leading order approximation $\psi^{(0)}(z)$,
the derivative $D^{\prime}_\mu(z)$ and the product $z D_\mu(z)$ both produce terms with $D_{\mu\pm 1}(z)$,
then to balance the equation the subleading order wavefunction $\psi^{(1)}(z)$ must contain terms with $D_{\mu}(z), D_{\mu\pm 2}(z), D_{\mu\pm 4}(z)$;
one can repeat this argument for higher order corrections, this explains the summation in the formula (\ref{PsiCSExpansion}).

The coefficients for the expansion of wavefunction display a general pattern as follows,
\begin{equation}
S_{\ell,2m}(\mu, A)=(-1)^m C_{\ell,2m}(\mu, -A).
\label{CoefficientPattern}
\end{equation}
One can always set $C_{0,0}=S_{0,0}=1$.
Moreover, since functions $D_n(z)$ form a complete orthonormal basis,
as usual in stationary perturbation theory higher order perturbations $\psi_{C/S}^{(\ell)}(z)$ with $\ell\geqslant 1$ are orthogonal to the leading order wavefunction $D_\mu(z)$,
therefore one could set $C_{\ell, 0}=S_{\ell, 0}=0$ for $\ell\geqslant 1$.
The first few $C_{\ell,m}$ up to order $B^{-1}$ are given by
\begin{align}
&C_{1,4}=-\frac{1}{16},\quad C_{1,2}=-\frac{1}{4},\quad C_{1,-2}=-\frac{1}{4}(\mu-1),\quad C_{1,-4}=\frac{\mu!}{16(\mu-2)!},   \nonumber\\
&C_{2,8}=\frac{1}{512},\quad C_{2,6}=\frac{1}{64},\quad C_{2,4}=-\frac{1}{16}(\mu+2),\quad C_{2,2}=\frac{\mu^2-25\mu-36-16A}{64},\nonumber\\
&C_{2,-2}=-\frac{\mu(\mu-1)(\mu^2+27\mu-10-16A)}{64},\quad C_{2,-4}=\frac{(\mu-1)\mu!}{16(\mu-4)!},\nonumber\\
&C_{2,-6}=-\frac{\mu!}{64(\mu-6)!},\quad C_{2,-8}=-\frac{\mu!}{512(\mu-8)!},\quad \cdots,
\end{align}
where $\mu!$ etc. are the factorial of integers.

The constants $C_0$ and $S_0$ are determined by normalization, in a similar way as Sips' treatment of Mathieu function \cite{Sips1949}.
For a deep well, the wavefunctions are exponentially suppressed beyond the classical turning points,
we can normalize wavefunctions in the internal between $-\pi/2$ and $\pi/2$. For $\psi_{C}(z)$ the normalization condition is
\begin{equation}
\int_{-\pi/2}^{\pi/2} \psi^2_{C}(z)d\phi=\int_{-\sqrt{2}B^{1/4}}^{\sqrt{2}B^{1/4}} \frac{\psi^2_{C}(z)}{\sqrt{2B^{1/2}-z^2}}dz=1.
\end{equation}
To compute the integral, we need to substitute the expression of $\psi_{C}(z)$ and expand the whole integrand as a series in large $B$,
then repeatedly use the first identity given in (\ref{IdentitiesParabolic}) of Appendix \ref{ParabolicCylinderFunction} to express the integrand in polynomials with products $D_{\mu}(z)D_{\mu^\prime}(z)$.
The integration interval for the variable $z$ can be extended from $-\infty$ to $\infty$,
because for $z^2>2B^{1/2}\gg 1$ the wavefunction is exponentially small so the extension of integration interval only introduces an exponentially small contribution.
In this way, one can compute the constant $C_0$ and $S_0$ as
\begin{align}
C_0&=\left(\frac{B^{1/4}}{\pi^{1/2}\mu!}\right)^{1/2}\left(1+\frac{2\mu+1}{B^{1/2}}+\frac{\mu^4+2\mu^3+263\mu^2+262\mu+108}{512B}+\cdots\right)^{-1/2},\nonumber\\
S_0&=\left(\frac{B^{1/4}}{\pi^{1/2}\mu!}\right)^{1/2}\left(1-\frac{2\mu+1}{B^{1/2}}+\frac{\mu^4+2\mu^3-121\mu^2-122\mu-84}{512B}+\cdots\right)^{-1/2},
\label{NormalizationConstSip}
\end{align}
up to the order given above it is the same as that for the Sips-Meixner solution of Mathieu functions \cite{NISTDLMF}, the corrections caused by $A$ appear in higher order terms.

Parity of the wavefunctions (\ref{wavefunction0S}) is deduced from parity of the parabolic cylinder function,
\begin{equation}
\psi_{C/S}(-\phi)=(-1)^\mu\psi_{C/S}(\phi),
\end{equation}
The Sips-Meixner expansion of Mathieu function $ce_\mu(\phi)$ and $se_{\mu+1}(\phi)$ are obtained from $\psi_{C}(\pi/2-\phi)$ and $\psi_{S}(\pi/2-\phi)$ in the limit $A\to 0$, respectively.

One can take the harmonic oscillator limit for the solution obtained in this section.
Set $\phi=l x$ and take the limit, $l\to 0$, $B\to \infty$, with $\mathscr{E} l^2=\varepsilon$ and $Bl^4=\alpha^4$ and the parameter $A$ fixed.
Then the equation becomes $\psi^{\,\prime\prime}(x)+(\varepsilon-\alpha^4x^2)\psi(x)=0$.
Only one term in the series expansion of eigenvalue (\ref{eigenvalue0}) survives in the limit, $\mathscr{E} l^2\to \varepsilon=2\alpha^2\tilde{\mu}$;
similarly, only the leading component in the series expansion of wavefunctions (\ref{wavefunction0S}) survives,
$\psi(x)\sim\psi_{C/S}^{(0)}(\sqrt{2}\alpha x)=D_\mu(\sqrt{2}\alpha x)=2^{-\mu/2}\exp(-\alpha^2x^2/2)H_\mu(\alpha x)$, where $H_\mu(\alpha x)$ is the Hermite polynomial.

\section{Oscillatory states at the saddle position $\phi_s=\pi$}\label{OscillatoryStatesAtPhiPi}

The oscillatory states at the saddle position $\phi_s=\pi$ are similar to that at $\phi_s=0$, computation of the spectrum is in parallel with the procedure in section \ref{WavefunctionInsideBarrier0}.
In fact, the integrand and wavefunctions for states at $\phi_s=\pi$ are obtained from the corresponding quantities for states at $\phi_s=0$ given in (\ref{integrands0}) and (\ref{wavefunction0Gpre}) by the following change of variables,
\begin{equation}
\phi\to \pi+\phi,\quad A\to -A,
\end{equation}
this operation keeps the potential (\ref{effectivepotential}) invariant. The wavefunctions are
\begin{align}
\widehat{\psi}_\pm(\phi)=&C_\pm\exp\bigg\{\mp B^{1/2}\cos\phi-\frac{1}{2}\ln\sin\phi\pm \frac{1}{2^4B^{1/2}}\Big[(8\mathcal{E}+1)\ln\cot\frac{\phi}{2}+\frac{3\cos\phi}{\sin^2\phi}\nonumber\\
& -8A\ln\sin\phi\Big]+\frac{1}{2^4B}\Big[\frac{3}{\sin^4\phi}-\frac{1-4\mathcal{E}-4A\cos\phi}{\sin^2\phi}\Big]+\cdots \bigg\},
\end{align}
they have a singularity at $\phi=\pi$, so are applicable in the regions away from this saddle position.
The wavefunction $\widehat{\psi}_+(\phi)$ has the correct decaying behavior in the barrier region near $\phi_s=\pi$, so it shall be used to join with wavefunction inside the well.

\begin{figure}[hbt]
\begin{minipage}[t]{0.5\linewidth}
\centering
\includegraphics[width=6cm]{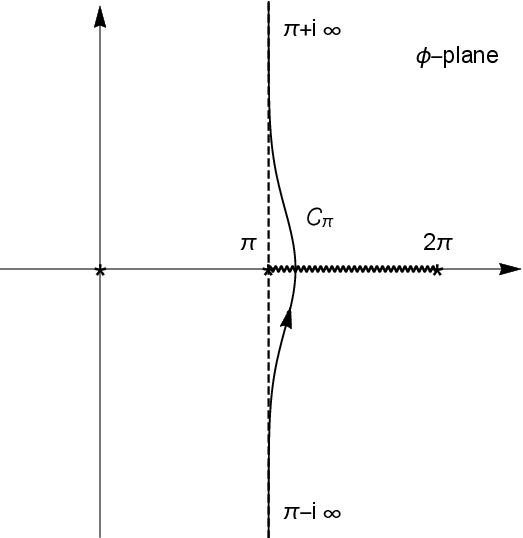}
\caption{Contour $\mathcal{C}_\pi$  and the branch cut\protect\\ for $\ln\sin\phi$ at $\phi_s=\pi$.} \label{PeriodContour4PhiPi}
\end{minipage}%
\begin{minipage}[t]{0.5\linewidth}
\centering
\includegraphics[width=6cm]{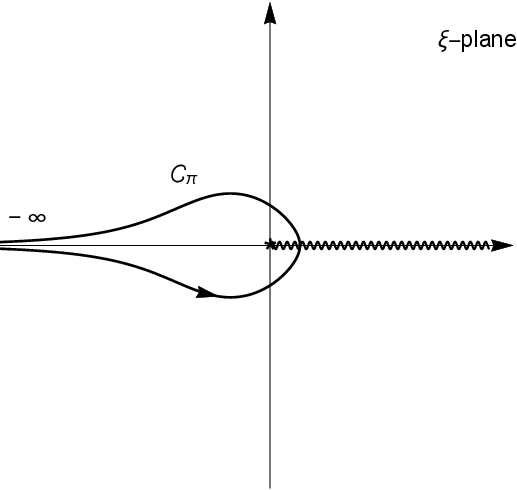}
\caption{Contour $\mathcal{C}_\pi$  in the $\xi$-coordinate and branch cut of $\ln\xi$.} \label{PeriodContour4XiPi}
\end{minipage}
\end{figure}

A quantum number for the oscillatory state can be introduced, we choose a contour $\mathcal{C}_\pi$ in the complex plane that passes close by the saddle $\phi_s=\pi$,
as shown in Fig. \ref{PeriodContour4PhiPi}, and define the corresponding index for $v_+(\phi)$ by
\begin{equation}
\mu=\frac{1}{i\pi}\int_{\mathcal{C}_\pi} v_+(\phi)d\phi,
\end{equation}
by the same reason as in section \ref{WavefunctionInsideBarrier0}, for $v_-(\phi)$ we have
\begin{equation}
-(\mu+1)=\frac{1}{i\pi}\int_{\mathcal{C}_\pi} v_-(\phi)d\phi.
\end{equation}
so that $\widehat{\psi}_\pm(\phi)$ are two independent wavefunctions associated to the same eigenvalue given by (\ref{eigenvaluePi}).
The integral is computed following the same method in section \ref{OscillatoryStatesAtPhi0},
using $\xi=\sin^2\phi$ the contour $\mathcal{C}_\pi$ in the $\phi$-pane is mapped to the contour in the $\xi$-plane as in Fig. \ref{PeriodContour4XiPi},
then formulae (\ref{integralformulae}) can be applied. The eigenvalue is given  by
\begin{equation}
\mathcal{E}_\pi=A+2B^{1/2}\tilde{\mu}-\frac{1}{8}(4\tilde{\mu}^2+1)-\frac{1}{32B^{1/2}}(4\tilde{\mu}^3+3\tilde{\mu}+16A\tilde{\mu})+\cdots,
\label{eigenvaluePi}
\end{equation}
where $\tilde{\mu}=\mu+1/2$ and $\mu$ is also a natural number.
The angular frequency of this well is $\Omega_\pi=2(B-A/2)^{1/2}$, lower than the angular frequency of the well at $\phi_s=0$.
The eigenfunctions become
\begin{align}
\widehat{\psi}_{\pi\pm}(\phi)=&C_{\pi\pm}\exp\bigg\{\mp B^{1/2}\cos\phi-\frac{1}{2}\ln\sin\phi\pm \tilde{\mu}\ln\cot\frac{\phi}{2}\pm\frac{1}{2^4B^{1/2}}\Big[\frac{\pm8\tilde{\mu}+(3+4\tilde{\mu}^2)\cos\phi}{\sin^2\phi} \nonumber\\ &-8A\ln\sin^2\frac{\phi}{2}\Big]+\frac{1}{2^6B}\Big[\frac{12+32\tilde{\mu}^2\pm(38\tilde{\mu}+8\tilde{\mu}^3)\cos\phi}{\sin^4\phi} \nonumber\\ &-\frac{(3+4\tilde{\mu}^2)(\mp\tilde{\mu}\cos\phi+2)-32A(1\pm\tilde{\mu})\sin^2\frac{\phi}{2}}{\sin^2\phi}\Big]+\cdots \Big\}.
\label{wavefunctionPiG}
\end{align}

Using the canonical coordinate defined in (\ref{CanoniCoordDef}), so that $\phi\to\phi+\mathcal{C}_\pi$ corresponds to $\rho\to \rho+i\pi$,
the eigenfunctions can be rewritten as quasi-periodic functions,
\begin{equation}
\widehat{\psi}_\pm(\rho)=\sqrt{\cosh\rho}\exp(\mp\widetilde{\mu}\rho)\hat{g}_\pm(\rho),
\end{equation}
where $\hat{g}_\pm(\rho)$ are functions of periodic $i\pi$, and the monodromy relation is
\begin{equation}
\widehat{\psi}_\pm(\rho+i\pi)=\sqrt{-1}\exp(\mp i\pi\widetilde{\mu})\widehat{\psi}_\pm(\rho)=\left( \begin{matrix}
e^{-i\pi(\mu+1)}  \cr e^{i\pi\mu}
\end{matrix}\right)\widehat{\psi}_\pm(\rho).
\label{monodromyCPicanoni}
\end{equation}

Inside the potential well at the saddle $\phi_s=\pi$, the wavefunction can be expanded by the parabolic cylinder function, too.
Using the same variable $z=\sqrt{2}B^{1/4}\sin\phi$, the wavefunction now satisfies the equation
\begin{equation}
(2B^{1/2}-z^2)\widehat{\psi}^{\,\prime\prime}(z)-z\widehat{\psi}^{\;\prime}(z)+\left(\mathscr{E}-A\sqrt{1-\frac{z^2}{2B^{1/2}}}-\frac{1}{2}B^{1/2}z^2\right)\widehat{\psi}(z)=0,
\label{SchrodEqZ2}
\end{equation}
with eigenvalue $\mathcal{E}$ given by (\ref{eigenvaluePi}).
The solutions also take the form of Sips-Meixner expansion,
\begin{align}
\widehat{\psi}_{C}(z)&=C_\pi \sum_{\ell=0}^\infty\frac{\widehat{\psi}_{C}^{(\ell)}(z)}{(2B^{1/2})^\ell},\quad \mbox{with}\quad \widehat{\psi}_{C}^{(\ell)}(z)=\sum_{m=-2\ell}^{2\ell}\widehat{C}_{\ell,2m}D_{\mu+2m}(z),\nonumber\\
\widehat{\psi}_{S}(z)&=S_\pi \cos\phi\sum_{\ell=0}^\infty\frac{\widehat{\psi}_{S}^{(\ell)}(z)}{(2B^{1/2})^\ell},\quad \mbox{with}\quad \widehat{\psi}_{S}^{(\ell)}(z)=\sum_{m=-2\ell}^{2\ell}\widehat{S}_{\ell,2m}D_{\mu+2m}(z),
\label{wavefunctionPiS}
\end{align}
then the coefficients $\widehat{C}_{\ell,2m}\equiv \widehat{C}_{\ell,2m}(\mu, A)$ and $\widehat{S}_{\ell,2m}\equiv \widehat{S}_{\ell,2m}(\mu, A)$ are solved, they are give by the following relations,
\begin{align}
&\widehat{C}_{\ell,2m}(\mu, A)=(-1)^m C_{\ell,2m}(\mu, -A),\nonumber\\
&\widehat{S}_{\ell,2m}(\mu, A)=(-1)^{m+1} S_{\ell,2m}(\mu, -A),
\end{align}
where $C_{\ell,2m}(\mu, A)$ and $S_{\ell,2m}(\mu, A)$ are coefficients obtained in section \ref{WavefunctionInsideWell0}.
The normalization constants $C_\pi$ and $S_\pi$ are computed following the same procedure as in section \ref{WavefunctionInsideWell0},
up to the first three orders the expressions are the same as that in (\ref{NormalizationConstSip}).

Now, we can summarize the perturbative spectral solutions obtained for oscillatory states.
Our method only leads to the eigensolution as piecewise functions, each of the wavefunctions applies in a portion of the space, despite they are formally periodic $2\pi$ function; they should be glued together to give the complete wavefunction, a matching condition should be applied in the overlapping region.
In the table \ref{SummaryPertSolu}, the formulae of eigenvalues and eigenfunctions are assigned to the  corresponding regions where they apply.  In this picture, a low energy quantum state is generally the linear combination of localized states in the wells around $\phi_s=0$ and around $\phi_s=\pi$.
That means in principle there is interference effect due to the overlap of the two localized wavefunctions;
but we show in the next section the overlap of wavefunctions is exponentially small when the wells are deep and asymmetric.

\begin{table}
\begin{center}
\renewcommand{\arraystretch}{1.00}
\begin{tabular}{ | c | c | c | c| c | }
\hline
\multicolumn{1}{|c|}{\multirow{2}{*}{solution$\diagdown$region}} & \multicolumn{2}{c|}{ states localize around $\phi_s=0$} & \multicolumn{2}{c|}{states  localize around $\phi_s=\pi$}\\ \cline{2-5}
 & \makecell{well \\ $\phi_t\lesssim\phi\lesssim\phi_t$ } &  \makecell{barriers \\ $\phi_t\lesssim|\phi|\lesssim\pi-\phi_t$ } & \makecell{well \\ $-\phi_t\lesssim\phi-\pi\lesssim\phi_t$}  & \makecell{barriers \\  $\phi_t\lesssim|\phi|\lesssim\pi-\phi_t$ } \\ \hline
eigenvalue & $\mathcal{E}_0$ in (\ref{eigenvalue0}) & (\ref{eigenvalue0}) &  $\mathcal{E}_\pi$ in (\ref{eigenvaluePi})  & (\ref{eigenvaluePi}) \\ \hline
eigenfunction & $\psi_{C/S}$ in  (\ref{wavefunction0S}) & $\psi_{0+}(\phi)$ in (\ref{wavefunction0G}) & $\widehat{\psi}_{C/S}$ in  (\ref{wavefunctionPiS}) &  $\widehat{\psi}_{\pi+}(\phi)$ in (\ref{wavefunctionPiG}) \\
\hline
\end{tabular}
\caption{Summary of perturbative eigensolution of oscillatory states.}
\label{SummaryPertSolu}
\end{center}
\end{table}

\section{Quantum tunneling}\label{NonperturbativeTunneling}

The tunneling phenomenon is characteristic for quantum system with multiple vacua \cite{ColemanSymmetry}.
Vacuum tunneling in quantum mechanics problem with trigonometric potential and symmetric double-well potential have been extensively discussed because the physical effects,
including energy bands and energy splitting, particle hopping, are observable in solids, molecules, quantum dots, Josephson junction, etc.
The effective potential (\ref{effectivepotential}) bears new features, the two vacua are asymmetric, and $\phi$ is the angle coordinate of a circle,
that means it is an {\em asymmetric double-well potential on a circle}.
Quantum mechanics with asymmetric double-well potential on the line has been discussed for example in \cite{Song2015, WeinerTse1981, Dekker1987, Song2008, Rastelli2012};
but nonperturbative quantum mechanics for the asymmetric double-well potential on a circle is rarely studied.

Quantum tunneling of asymmetric double-well depends sensitively on the shape of potential. The oscillation frequency of states in the wells is $2B^{1/2}$,
when the potential is weakly asymmetric, $A\ll 2B^{1/2}$, perturbative energy levels in the two wells are nearly degenerate.
Since the condition $A\ll B^{1/2}\widetilde{\mu}\sim B^{1/2}$ is assumed for states with small index $\tilde{\mu}$, here we discuss low-lying states in weakly asymmetric double-wells.

Let $|\mu\rangle_0$ denotes the perturbative state with quantum number $\mu$ at $\phi_s=0$ and similarly $|\mu\rangle_\pi$ for the perturbative state at $\phi_s=\pi$;
they have an energy difference of order $2A$, both are widely separated from adjacent energy levels.
The problem can be recast as a two-level quantum system with weak quantum tunneling.
The two perturbative states get mixed,
let $|\mu\rangle_-$ and $|\mu\rangle_+$ denote the resulting nonperturbative eigenstates,
\begin{equation}
 |\mu\rangle_-=\cos\theta|\mu\rangle_0+\sin\theta|\mu\rangle_\pi, \quad |\mu\rangle_+=-\sin\theta|\mu\rangle_0+\cos\theta|\mu\rangle_\pi.
\end{equation}
The nonperturbative eigenvalues $\mathcal{E}_\pm(\mu)$ can be computed from diagonalization of the Hamiltonian
\begin{equation}
H=\left( \begin{matrix}
\mathcal{E}_0(\mu) &  -\gamma \cr -\gamma & \mathcal{E}_\pi(\mu)
\end{matrix}\right),
\label{2StateHamiltonian}
\end{equation}
where $\mathcal{E}_0(\mu)$ is given by (\ref{eigenvalue0}), $\mathcal{E}_\pi(\mu)$ is given by (\ref{eigenvaluePi}),
and $\gamma$ denotes the well-to-well tunneling effect.
The tunneling effect is exponentially small for deep wells, using the WKB approximation it is given by
\begin{equation}
\gamma\approx \frac{2g_\mu\sqrt{\Omega_0\Omega_\pi}}{\pi}\exp(-S),
\label{TunnelingMatrix}
\end{equation}
where we have used the result for asymmetric double-well potential obtained in \cite{Song2015}.
As far as tunneling through one barrier is considered, the prefactor is the same;
the factor $2$ accounts for the fact that the tunneling between the two wells on a circle could occur either clockwise or anticlockwise \cite{GolovinskiDubinkin2021}.
The function $g_\mu=\sqrt{2\pi}\tilde{\mu}^{\tilde{\mu}}\exp(-\tilde{\mu})/\mu!$ is the Furry factor \cite{Furry1947},
used to achieve a better WKB approximation for wavefunction with small quantum number \cite{Bhattacharya1985, MullerKirstenQUANT};
the angular dependent factor is
\begin{equation}
\sqrt{\Omega_0\Omega_\pi}=2B^{1/2}\left(1-\frac{A^2}{4B^2}\right)^{1/4}\approx 2B^{1/2}-\frac{A^2}{8B^{3/2}}+\cdots.
\end{equation}
The potential is symmetric w.r.t the two saddle positions, so the tunneling action in the exponential factor for the left and right barriers are the same.
For the barrier in the region $0\leqslant\phi\leqslant \pi$, the tunneling action is computed by
\begin{equation}
S=\int_{\phi_t}^{\pi-\phi_t}\sqrt{u(\phi)-\mathcal{E}(\mu)}d\phi.
\end{equation}
Because states $|\mu\rangle_0$ and  $|\mu\rangle_\pi$ have slightly mismatched energies,  their tunneling actions are not the same.
For the state $|\mu\rangle_0$ the turning points are determined by $u(\phi_t)=\mathcal{E}_0(\mu)$, the tunneling action is
\begin{equation}
S_+=2B^{1/2}-\frac{9\tilde{\mu}+2\tilde{\mu}\ln(4B/\tilde{\mu}^2)}{4}+\mathcal{O}(B^{-1/2}).
\label{TunnelingActions}
\end{equation}
Similarly, for the state $|\mu\rangle_\pi$ the tunneling action is smaller, $S_-\approx S_+-A[3+2\ln (4B/\tilde{\mu}^2)]/4B^{1/2}$.
Since the WKB computation gives only approximate estimate of the tunneling coupling for low-lying states,
we ignore the small difference and set $S\approx 2B^{1/2}$ to keep the Hamiltonian (\ref{2StateHamiltonian}) Hermitian.

The eigenvalues of Hamiltonian are
\begin{equation}
\mathcal{E}_\pm(\mu)=\frac{1}{2}[\mathcal{E}_\pi(\mu)+\mathcal{E}_0(\mu)]\pm\frac{1}{2}\Delta(\mu),
\end{equation}
where $\Delta(\mu)$ is the broadened energy bias,
\begin{equation}
\Delta(\mu)=\sqrt{[\mathcal{E}_\pi(\mu)-\mathcal{E}_0(\mu)]^2+4\gamma^2}.
\end{equation}
The states mixing angle is given by
\begin{equation}
\tan2\theta=\frac{2\gamma}{\mathcal{E}_\pi(\mu)-\mathcal{E}_0(\mu)}\approx \frac{\gamma}{A}(1+\frac{\tilde{\mu}}{2B^{1/2}}+\cdots).
\label{TunnelingMixing}
\end{equation}
For $A\ne 0$ and $\gamma$ exponentially small, the mixing is exponentially suppressed.
In the unbiased limit $A\to 0$, states are maximally mixed,
one gets the well-known result of energy splitting for symmetric double-well, the antisymmetric state $|\mu\rangle_+$ has a higher energy.
The formula (\ref{TunnelingMixing}) leads to the following qualitative conclusion.
The energy bias suppresses quantum tunneling, this is a general feature of asymmetric double-wells \cite{NietoStrottman1985, BurrowsFeldmann1998}.
The states of higher energy level have a larger tunneling probability than the states of lower energy level.
This is good for comparing the results obtained in previous sections and future experiments: low-lying quantum states localize in the two wells are well separated,
the perturbative eigensolutions remain accurate;  if the dissipation is reduced to weak enough,
a state trapped in a well remains intact for sufficiently long lifetime for experimentation.

We here consider tunneling effects only in a simplified context, in a more careful consideration new phenomenon would occur.
For example, when the asymmetry parameter increases larger to $A\approx B^{1/2}$, which of course lies outside the perturbative region of this paper,
fine tuned to match energies of states $|\mu+1\rangle_0$ and $|\mu\rangle_\pi$, resonant tunneling would occur \cite{Song2008, NietoStrottman1985}.
Similar to the study on symmetric double-well potential on a circle \cite{MantonSamols1988, Khlebnikov1991, Liang1992},
one can study quantum tunneling using periodic instanton for the asymmetric case.
It is also interesting to consider the connection between the large order behavior of perturbative spectrum and the nonperturbative tunneling effects, a phenomenon explained in e.g. chapter 20 of \cite{MullerKirstenQUANT}.

\section{Conclusion}

In this paper we study the spectral aspect of oscillatory states in a quantum mechanical system with Kapitza pendulum effective potential.
A generalisation of Floquet theorem, originally applied to similar spectral problem with elliptic potentials, is used to compute the series expansion of eigenvalue and wavefunction.
The wavefunction of confined states have simple zeros in the classically allowed region,
then the quantization condition can be expressed as contour integral for the integrand of wavefunction along an imaginary trajectory.
The quantum number takes value in nature number since it equals the number of nodes of wavefunction;
the factor $1/2$ of zero point energy appears naturally from computation of contour integral.

The complex nature of quantum mechanics is usually understood that quantities such as operators, wavefunction and phase factor are complex, but complex coordinates are rarely considered.
Nevertheless, it sometimes happens that extending coordinates to complex domain provides a clearer explanation,
here is the example of quantum harmonic oscillator,
the drastic change of wavefunction from oscillatory behavior to decaying behavior near the turning point is explained by the Stokes phenomenon of Airy function in the complex plane.
The study in this paper suggests it is useful to consider quantum mechanical problem in complex coordinate, it provides not only a convenient method to compute,
but also new view on some basic notions such as the quantization.

\appendix

\section{Floquet theorem for elliptic potential}\label{Appendix1}

One may wonder if the ``{\em imaginary period of infinite length}" in the $\phi$-plane of section \ref{OscillatoryStatesAtPhi0} really make sense, after all,
the trigonometric function only has a real period $2\pi$. We here explain how the computation based on ``infinitely extended period" evolves from the spectral problem of potential with elliptic functions where the periods are finite.

The trigonometric functions can be obtained from the Jacobian elliptic functions by taking the limit of vanishing elliptic modulus, $k\to 0$,
for example, $\varinjlim\sn(\phi,k^2)=\sin\phi$ and $\varinjlim\cn(\phi,k^2)=\cos\phi$.
For brevity, the notations $\sn\,\phi\equiv \sn(\phi,k^2)$ and $\cn\,\phi\equiv \cn(\phi,k^2)$ are used.
Therefore, the trigonometric potential (\ref{effectivepotential}) is the limit of an elliptic potential,
\begin{equation}
u(\phi)=-A\cn\,\phi+B\sn^2\phi. \label{ellipticpotential}
\end{equation}
The elliptic potential (\ref{ellipticpotential}) has two independent periods $4\mathbf{K}$ and $2\mathbf{K}+2i\mathbf{K}^\prime$,
given by complete elliptic integral of the first kind, $\mathbf{K}\equiv \mathbf{K}(k^2)$ and $\mathbf{K}^\prime\equiv \mathbf{K}(1-k^2)$.
The imaginary period $4i\mathbf{K}^\prime$ is obtained by combination.
Elliptic functions are intrinsically complex, in the trigonometric limit only real period survives,
$4\mathbf{K}\to 2\pi$ while $4i\mathbf{K}^\prime\to i\infty$.
This is reason we say the potential (\ref{effectivepotential}) has the usual real period $2\pi$,
when extended to the complex domain it also has an imaginary period of infinite extent from $-i\infty$ to $+i\infty$,
this geometric picture  is useful to understand the discussion in section \ref{OscillatoryStatesAtPhi0}.

The eigensolutions for the Schr\"{o}dinger equation with elliptic Hill potentials are studied in \cite{wh1412, wh1608, wh1904},
using a generalisation of Floquet theorem applicable to all three periods of elliptic function.
A representative example is the spectral problem of the  Lam\'{e} potential $u(\phi)=B\sn^2\phi$ with periods $2\mathbf{K}$ and $2i\mathbf{K}^\prime$.
The small eigenvalue solution around a saddle point, say the one at $\phi_s=0$,
is given by a Floquet solution associated to the period $2i\mathbf{K}^\prime$.
The potential (\ref{ellipticpotential}) contains the term $-A\cn\,\phi$ which breaks the periodicity of the term $B\sn^2\phi$, nevertheless,
the Floquet theorem applies because with $B\gg A$ it can be considered as a perturbation of the Lam\'{e} potential.
The saddle positions of the potential (\ref{ellipticpotential}) are $\phi_s=0, \mathbf{K}, 2\mathbf{K}, 3\mathbf{K}$ and $\mathbf{K}+i\mathbf{K}^\prime, 2i\mathbf{K}^\prime$, etc.
In the trigonometric limit, saddle positions $\phi_s=0$ and $\phi_s=2\mathbf{K}$ become bottom of wells while $\phi_s=\mathbf{K}$ and $\phi_s=3\mathbf{K}$ become summit of barriers, the saddle positions with imaginary components are sent to infinity region.
As shown in previous study, for the small energy solution around the saddle at $\phi_s=0$,
the index $\mathcal{\mu}$ is computed from
\begin{equation}
\mu=\frac{1}{i\pi}\int_{-i\mathbf{K}^\prime}^{i\mathbf{K}^\prime} v_+(\phi)d\phi,
\label{periodintegral0pElliptic}
\end{equation}
here the integrand $v_+(\phi)=-B^{1/2}\sn\,\phi-\frac{1}{2}\partial_\phi\ln\sn\,\phi+\cdots$ is the large-$B$ series solution of the equation (\ref{eq4integrand}) with the elliptic potential (\ref{ellipticpotential}),
the contour $\mathcal{C}_0$ is a curve of finite length from $-i\mathbf{K}^\prime$ to $i\mathbf{K}^\prime$ representing the period of the dominant potential term $B\sn^2\phi$. The following four kinds of integrals arise in the formula (\ref{periodintegral0pElliptic}),
\begin{equation}
\int_{-i\mathbf{K}^\prime}^{i\mathbf{K}^\prime}\frac{1}{\sn^{2n-1}\phi}d\phi,\quad \int_{-i\mathbf{K}^\prime}^{i\mathbf{K}^\prime}\frac{\cn\,\phi}{\sn^{2n+1}\phi}d\phi,\quad \int_{-i\mathbf{K}^\prime}^{i\mathbf{K}^\prime}\frac{\dn\,\phi}{\sn^{2n+1}\phi}d\phi,\quad
\int_{-i\mathbf{K}^\prime}^{i\mathbf{K}^\prime}\frac{\cn\,\phi\dn\,\phi}{\sn^{2n+1}\phi}d\phi
\end{equation}
with integer $n\geqslant 0$.
Using the map $\sn^2\phi=\xi$, the integrals over period $2i\mathbf{K}^\prime$ in the $\phi$-plane are transformed to integrals along a closed contour in the $\xi$-plane.
Among the fourth kind of integrals, only the integral with $n=0$ leads to nonzero value $i\pi$,
which is responsible for the ``zero point energy" as in the trigonometric case; other integrals with $n\geqslant 1$ vanish.
The remaining three kinds of integrals become
\begin{equation}
\frac{1}{2}\oint_{\mathcal{C}_0}\frac{d\xi}{\xi^n\sqrt{(1-\xi)(1-k^2\xi)}},\qquad \frac{1}{2}\oint_{\mathcal{C}_0}\frac{d\xi}{\xi^{n+1}\sqrt{1-k^2\xi}},\qquad
\frac{1}{2}\oint_{\mathcal{C}_0}\frac{d\xi}{\xi^{n+1}\sqrt{1-\xi}},
\end{equation}
expanding the integrand near $\xi=0$, the integrals are evaluated by the residue theorem,
similar to the computation in \cite{wh1412, wh1608, wh1904}.
The integral (\ref{periodintegral0pElliptic}) gives the series expansion of $\mu(\mathcal{E})$,
from which one obtains the eigenvalue expansion,
\begin{equation}
\mathcal{E}=-A+2B^{1/2}\tilde{\mu}-\frac{1}{8}(1+k^2)(4\tilde{\mu}^2+1)-\frac{1}{32B^{1/2}}[(1+k^2)^2(4\tilde{\mu}^3+3\tilde{\mu})-4k^2(4\tilde{\mu}^3+5\tilde{\mu})-16A\tilde{\mu}]+\cdots,
\label{eigenvalue0Elliptic}
\end{equation}
with $\tilde{\mu}=\mu+1/2$.
There is another integrand solution $v_-(\phi)$, obtained from $v_+(\phi)$ by the inversion $B^{1/2}\to -B^{1/2}$,
as in section \ref{WavefunctionInsideBarrier0}, by the relation
\begin{equation}
-(\mu+1)=\frac{1}{i\pi}\int_{-i\mathbf{K}^\prime}^{i\mathbf{K}^\prime} v_-(\phi)d\phi,
\label{periodintegral0nElliptic}
\end{equation}
it leads to the same eigenvalue given by (\ref{eigenvalue0Elliptic}).
However, for the case of elliptic potential, there is no argument to ensure the index $\mu$ an integer,
unless some particular boundary conditions of wavefunctions are assigned.
In the limit $A\to 0$, the expression (\ref{eigenvalue0Elliptic}) reduces to the eigenvalue of Lam\'{e} potential;
and in the limit $k\to 0$, it reduces to the eigenvalue of Kapitza potential (\ref{eigenvalue0}).

The corresponding wavefunctions are also piecewise,
in the region near $\phi=\mathbf{K}$ they are represented by functions obtained from indefinite integrals of $v_\pm(\phi)$,
in the region near $\phi=0$ they are represent by functions expanded in parabolic cylinder function.
For the small energy solution around the saddle at $\phi_s=2\mathbf{K}$,
the eigensolution is obtained from that at $\phi_s=0$ by the change of variables $\phi\to \phi+2\mathbf{K}, A\to -A$,
similar to the argument in section \ref{OscillatoryStatesAtPhiPi}.
The details can be worked out following sections \ref{OscillatoryStatesAtPhi0} and \ref{OscillatoryStatesAtPhiPi}, but to avoid departure too far from the main theme, we halt the discussion on the elliptic potential (\ref{ellipticpotential}).

In the trigonometric limit, the contour becomes infinitely extended.
But the difference is not that severe as it looks, the contour of period for elliptic potential is finite in the $\phi$-plane,
but the potential has poles, on the equienergy surface the contour is infinitely stretched;
there is no pole for trigonometric potential, this compensates the infiniteness of the contour.
However, for the equation with elliptic potential not everything goes in parallel with that of trigonometric potential,
for example, there is no notion of classically allowed and forbidden regions,
and it is not clear how much it changes the distribution of the zeros of wavefunction in the complex plane.

\section{Asymptotic coincidence of eigenvalues}\label{EigenvalueCurveMathieu}

The Mathieu equation $\psi^{\,\prime\prime}(x)+(\lambda-2h\cos2x)\psi(x)=0$ is a prototype of ODE with periodic potential \cite{Arscott1964, mclachlan1947, Magnus-Winkler1966, Eastham1973, NISTDLMF},
for every $\lambda_\nu$ it has the even parity solution $ce_\nu(x)$ and the odd parity solution $se_\nu(x)$, where $\nu$ is the index.
Without imposing periodic boundary condition, the index $\nu$ in general takes value in complex number; however,
only solutions with real index are stable. For solutions with even integer index, $ce_{2n}(x)$ and $se_{2n}(x)$ are periodic $\pi$;
for solutions with odd integer index, $ce_{2n+1}(x)$ and $se_{2n+1}(x)$ are periodic $2\pi$.

\begin{figure}
\begin{center}
\includegraphics[width=6cm]{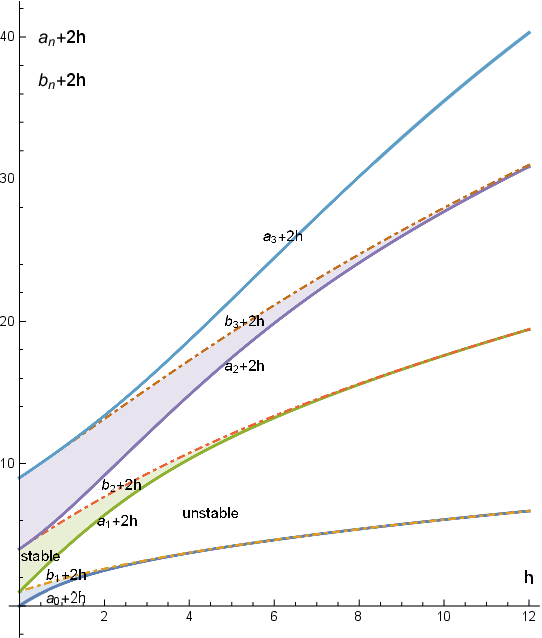}
\end{center}
\caption{Curves of $a_n(h)$ and $b_n(h)$ up to $n=3$, with bands and gaps.} \label{BandsGapsChartMathieu}
\end{figure}

The strength of the potential $h$ can be tuned, from weak coupling (small potential) to strong coupling (large potential).
Series solutions can be computed for both weak and strong coupling, numerical solutions for general coupling strength are implemented in computation tools.
An important property of the spectrum chart is the existence of stable regions (bands) and unstable regions (gaps),
the boundaries are curves of $a_n(h)$ and $b_n(h)$ with integer index $n=0,1,2,3\cdots$, as depicted in Fig. \ref{BandsGapsChartMathieu}.
One notices an interesting feature in the spectrum chart, that for small $h$, eigenvalues $a_n$ and $b_n$ with $n\geqslant 1$ asymptotically coincide in the limit $h\to 0$,
in fact when $n\gg1$ they have the same weak coupling series expansion,
\begin{equation}
\left( \begin{matrix}
a_n  \cr b_n
\end{matrix}\right)
=n^2+\frac{1}{2(n^2-1)}h^2+\frac{5n^2+7}{32(n^2-1)^3(n^2-4)}h^4+\mathcal{O}(h^6),
\end{equation}
they differ by perturbative terms of order $\mathcal{O}(h^n)$.
On the other end of large $h$, eigenvalues $a_n$ and $b_{n+1}$ with $n\geqslant 0$ asymptotically coincide, indeed they have the same strong coupling series expansion,
\begin{equation}
\left( \begin{matrix}
a_n  \cr b_{n+1}
\end{matrix}\right)
=-2h+4(n+\frac{1}{2})h^{1/2}-\frac{1}{8}[4(n+\frac{1}{2})^2+1]+\mathcal{O}(h^{-1/2}),
\end{equation}
they differ by nonperturbative terms of order $\mathcal{O}(e^{-4\sqrt{h}})$. A detailed discussion of the stability chart for Mathieu equation can be found in \cite{mclachlan1947}.

\begin{figure}
\begin{center}
\includegraphics[width=12cm]{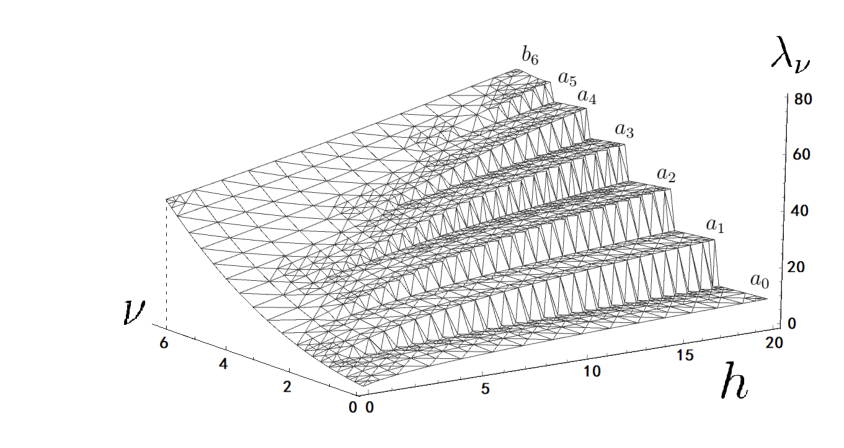}
\end{center}
\caption{Eigenvalue $\lambda_\nu(h)$ for $n<\nu<n+1$ approaches $a_n\simeq b_{n+1}$ for large $h$.} \label{MathieuEigenvalueNonintegerOrder}
\end{figure}

Another interesting feature of the spectrum chart is about the behavior of eigenvalue $\lambda_\nu$ with fractional index $n<\nu<n+1$.
For the case of weak coupling, the eigenvalue increases smoothly with the index as $\lambda_\nu\sim \nu^2$; when coupling grows strong,
the eigenvalue $\lambda_\nu$ asymptotically coincides the eigenvalues $a_n\approx b_{n+1}$ with integer index.
The transition of asymptotic behavior of $\lambda_\nu$ with increasing $h$ is shown in the three dimensional spectrum chart Fig. \ref{MathieuEigenvalueNonintegerOrder}.

For the equation (\ref{SchrodEqPhi}) with potential (\ref{effectivepotential}), the full stability chart is three dimensional because there are two couplings $A$ and $B$,
it could be quite complicated. It is possible to obtain a precise numerical study on the full stability chart, by applying the method of Hill's infinite determinant to Whittaker-Hill equation.
Only a small portion of the parameter space is physically relevant to the quantum Kapitza pendulum. Especially for the strong coupling solution discussed in section \ref{OscillatoryStatesAtPhi0},
the relevant region of parameter space is a narrow volume with small $A$, large positive $B$ and $\mathcal{E}\sim 2n B^{1/2}$,
the asymptotic behavior of stability chart in this region should be close that of Mathieu equation.

\section{Parabolic cylinder function}\label{ParabolicCylinderFunction}

The parabolic cylinder function $D_n(z)$ with integer $n$ satisfies Weber's equation
\begin{equation}
D^{\,\prime\prime}_n(z)+(n+\frac{1}{2}-\frac{z^2}{4})D_n(z)=0.
\end{equation}
Both derivative and product change the index $n$,
\begin{align}
& zD_n(z)=nD_{n-1}(z)+D_{n+1}(z), \nonumber\\
& D^{\,\prime}_n(z)=\frac{1}{2}nD_{n-1}(z)-\frac{1}{2}D_{n+1}(z).
\label{IdentitiesParabolic}
\end{align}
The parity is $D_n(-z)=(-1)^n D_n(z)$.
The function is bounded around the origin, $D_n(z)\sim \exp(-z^2/4)z^n$ for $|z|\gg |n|$.
It has an analytic continuation in the whole complex plane. The  orthogonality relation is
\begin{equation}
\int_{-\infty}^{\infty}D_m(z)D_n(z)dz=n!(2\pi)^{1/2}\delta_{mn}.
\end{equation}

\section*{Acknowledgments}

This work is supported by a grant from CWNU (No. 18Q068), a grant from Science \& Technology Department of Sichuan Province (No. 2021ZYD0032), a grant from Chinese Universities Scientific Fund (No. 2452018158).

\end{document}